\documentclass[12pt]{article}
\usepackage{amsmath,amsfonts}
\usepackage{hyperref}
\usepackage{graphicx}
\usepackage{xcolor}
\usepackage[numbers,sort&compress]{natbib}

\usepackage{amsthm}
\usepackage{amssymb}
\usepackage[matrix,arrow]{xy}
\usepackage{epsfig}
\usepackage{slashed}

\usepackage{braket}

\makeatletter\@addtoreset{equation}{section}\makeatother

\newcommand{\al}{\alpha}
\newcommand{\be}{\beta}
\newcommand{\te}{\theta}
\newcommand{\la}{\lambda}

\newcommand{\dr}{\partial_r}

\newcommand{\zb}{\bar{z}}

\newcommand{\onov}[1]{\frac{1}{#1}}
\newcommand{\mat}[1]{\left(\begin{matrix} #1 \end{matrix}\right)}

\newcommand{\lag}{\mathcal{L}}

\newcommand{\beq}{\begin{equation}}
\newcommand{\eeq}{\end{equation}}
\newcommand{\bea}{\begin{eqnarray}}
\newcommand{\eea}{\end{eqnarray}}
\newcommand{\half}{\frac{1}{2}}

\newcommand{\eq}[2][ ]{\begin{equation}\label{#1}{\begin{split}#2\end{split}}\end{equation}}
\newcommand{\eql}[2]{\begin{equation}\label{#1}{\begin{split}#2\end{split}}\end{equation}}

\newcommand{\hm}{\hat\mu}

\newcommand{\hx}{\hat{\xi}}

\newcommand{\cD}{{\mathcal D}}

\newcommand{\cN}{{\mathcal N}}

\newcommand{\cO}{{\mathcal O}}

\renewcommand{\title}[1]{\vbox{\center\LARGE{#1}}\vspace{5mm}}
\renewcommand{\author}[1]{\vbox{\center\large#1}\vspace{5mm}}

\begin{document}

\bibliographystyle{utphys}
	
\begin{titlepage}
		\vspace{8mm}
	\begin{center}
		\title{\bf Vortex-Strings in $\cN=2$ SQCD and Bulk-String Decoupling}
		\vspace{8mm}
		{\large \bf Efrat Gerchkovitz}\footnote{efrat.gerchkovitz@weizmann.ac.il}
		{\bf and}
		{\large \bf Avner Karasik}\footnote{avner.karasik@weizmann.ac.il}\\\vspace{2mm}{\large\textit{ Department of Particle Physics and Astrophysics,\\Weizmann Institute of Science, Rehovot 76100, Israel}}
	\end{center}

	\vspace{5mm}
\begin{abstract}
	We study vortex-strings in four-dimensional $\cN=2$ supersymmetric $SU(N_c)\times U(1)$ gauge theories with $N_f$ hypermultiplets in the fundamental representation of $SU(N_c)$ and general $U(1)$ charges. If $N_f>N_c$, the vacuum is not gapped and the low-energy theory contains both the vacuum massless excitations and the string zero-modes. The question we address in this work is whether the vacuum and the string moduli decouple at low energies, allowing a description of the low-energy dynamics in terms of a two-dimensional theory on the string worldsheet.   We find a simple condition  controlling the bulk-string coupling: If there exist two flavors such that the product of their $U(1)$ charge difference with the magnetic flux carried by the string configuration  is not an integer multiple of $2\pi$, the string has zero-modes that decay slower than $1/{r}$, where $r$ is the radial distance from the string core. These modes are coupled to the vacuum massless excitations even at low energies. If, however, all such products are integer multiples of $2\pi$, long-range modes of this type do not exist and the string moduli decouple from the bulk at low energies. This condition turns out to coincide with the condition of trivial Aharonov-Bohm phases for the particles in the spectrum. In addition to a derivation of the bulk-string decoupling criterion using classical analysis of the string zero-modes, we provide a non-perturbative derivation of the criterion, which uses supersymmetric localization techniques. 
\end{abstract}

\end{titlepage}

	\section{Introduction and Summary}
	
	One of the most important tools in the study of solitons is the moduli space approximation \cite{MANTON198254}, in which the low-energy dynamics in the soliton background is governed by the moduli space of solutions to the  soliton equations.
	
	If the only massless excitations are the soliton collective coordinates, the low-energy effective theory in the background of the soliton can be described as a sigma model living on the soliton worldvolume. The target space of the sigma model is the soliton moduli space, equipped with a metric induced by the underlying gauge theory.  This approach leads to interesting relations between theories of different dimensions.

	If, however,  the bulk theory is not gapped, the low-energy dynamics will be controlled by both the bulk excitations and the soliton excitations. In general, one expects that the worldvolume theory cannot be separated from the bulk.  This difficulty is avoided if the interactions between the bulk and the soliton massless excitations are suppressed  such that at energies below some mass scale, the bulk and the soliton decouple and the low-energy effective theory factorizes,
	\eql{factor}{S_{\text{eff}}=S_{\text{bulk}}+S_{\text{soliton}}\ .}
	For example, in the context of scattering of BPS monopoles, it has been argued that the moduli space approximation is applicable at low velocities, despite the fact that the bulk theory is not gapped \cite{Manton:1988bn}.

	In this work, we address the question of bulk-soliton decoupling in a particular class of BPS vortex-strings in four-dimensional $\cN=2$ supersymmetric gauge theories.
	Such strings have been extensively studied  in the case where the gauge group is $U(N_c)$ and the matter content consists of $N_f$ fundamental hypermultiplets. See, for example, \citep{Auzzi:2003fs,Hanany:2003hp,Hanany:2004ea,Shifman:2004dr,Shifman:2006kd,Eto:2005yh,Auzzi:2005gr,Eto:2007yv}, and the reviews \cite{Shifman:2007ce,Tong:2005un,Tong:2008qd,Eto:2006pg}. Let us start by reviewing this well-studied case. 
	
	When $N_f=N_c=N$ the low-energy effective theory in the minimal string background is given by a two-dimensional $\cN=(2,2)$ supersymmetric  ${\mathbb{CP}}^{N-1}$ non-linear sigma model, living on the worldsheet of the string. This result, which was derived in various methods, such as brane construction \citep{Hanany:2004ea,Hanany:2003hp} and an explicit expansion around the string solution \citep{Auzzi:2003fs,Shifman:2004dr,Hanany:2004ea},   provides an explanation for an earlier observation \cite{Dorey:1998yh}, relating the BPS spectra  and beta-functions of the four-dimensional $\cN=2$ gauge theory and the two-dimensional $\cN=(2,2)$ ${\mathbb{CP}}^{N-1}$ sigma model.  
	
	For $N_f>N_c$, a two-dimensional effective description of the low-energy dynamics around the string background was derived in similar methods \cite{Hanany:2004ea,Hanany:2003hp,Shifman:2006kd}.  In contrast to the $N_f=N_c$ case, the target space of the sigma model is non-compact when $N_f>N_c$. This is a result of the fact that the moduli space of the string has a  non-compact direction, related to the size of the string in the transverse plane  \citep{Vachaspati:1991dz}. The size zero-modes of the string will play a central role in this work.
	
	The results mentioned above have been generalized to strings with non-minimal winding numbers \cite{Auzzi:2005gr,Hanany:2003hp,Eto:2005yh,Hashimoto:2005hi,Eto:2006cx} and to other gauge groups \citep{Eto:2009zz, Ferretti:2007rp}.

	For $N_f=N_c$,   the worldsheet theory is assured to capture   the low-energy dynamics of the theory 
	since the bulk theory is gapped. The only massless excitations in the string background are the string moduli and the low-energy dynamics is therefore two-dimensional.  When  $N_f>N_c$, however, there are massless Goldstone excitations in the bulk. The question of whether factorization of the form of \eqref{factor} applies is crucial for the understanding of the low-energy dynamics in the background of the string. 
	To the best of our knowledge, this question has not been directly addressed  in the literature. Yet, the  two-dimensional description obtained in the moduli space approximation reproduces the correct BPS spectrum and beta-functions~\cite{Dorey:1999zk}. This agreement suggests that the bulk and the string degrees of freedom indeed decouple at low energies.  We will obtain this factorization as a special case of our results.   
	
	In this paper, we study BPS vortex-strings in four-dimensional $\cN=2$ supersymmetric gauge theories with gauge group $SU(N_c)\times U(1)$ and $N_f$ hypermultiplets in the fundamental representation of $SU(N_c)$. The charges of the hypermultiplets under the $U(1)$ factor of the gauge group will be denoted by $c_i\in\mathbb{Z}$, $i=1,..,N_f$.  In other words, we generalize the set-up of \citep{Auzzi:2003fs,Hanany:2003hp,Hanany:2004ea,Shifman:2004dr}, in which $c_i=1$ for all $i$, by allowing different $U(1)$ charges for different hypermultiplets. 
	
	The strings that we will be interested in are labeled by the winding number $K\in\mathbb{Z}$ and by a choice of $N_c$ out of the $N_f$ hypermultiplets that obtain a vacuum expectation value. The vacuum equations require that the sum of the $U(1)$ charges of these $N_c$ hypermultiplets will be non-vanishing. We denote this sum by $C$. 
	The magnetic flux carried by the string is $\Phi={2\pi K}/{C}$. Topologically, such fluxes are enabled by a breaking of the gauge group to a residual  $\mathbb{Z}_C$ subgroup  in the vacuum.  
	
	We focus on the $N_f>N_c$ case, for which the bulk theory is not gapped  and the question of bulk-string factorization arises. 
	Assuming an ansatz for the string solution and studying its size-modes and their interactions with the bulk massless modes, we obtain a criterion controlling their coupling to the bulk at low energies: if  $({c_i-c_j})\Phi\in2\pi\mathbb{Z}$ for all $i, j$, the mixing terms are suppressed at energies $E\ll m_W$, where $m_W$ denotes the mass of the W-bosons. In these cases, we have a factorization of the form (\ref{factor}). Note that the equal charge case of \citep{Auzzi:2003fs,Hanany:2003hp,Hanany:2004ea,Shifman:2004dr} satisfies this condition. 
	
	When the condition above is not satisfied, there exist long-range moduli that decay as $r^{-\be}\ ,\ 0<\be<1$, where $r$ is the radial coordinate on the transverse plane to the string. Long-range modes of this type are not very common in the literature.  As we will show, these long-range modes are mixed with the bulk modes even at low energies, and as a result, the low-energy effective theory does not factorize.  We thus find a criterion for the decoupling of the string from the bulk at low energies, 
	\begin{equation}
	\text{\bf Decoupling Criterion:}\;\;\;\;\;\;\;\frac{({c_i-c_j})\Phi}{2\pi}=\frac{({c_i-c_j})K}{C} \in\mathbb{Z} \;\;\text{for all}\; i,j\;. \label{decoupling criterion}
	\end{equation}
	Note that when $N_f=N_c$ there are no light fields in the bulk, and  the low-energy dynamics is described by the two-dimensional worldsheet theory, regardless of whether \eqref{decoupling criterion} is satisfied or not.
	
	Using supersymmetric localization techniques, we obtain a non-perturbative evidence for the decoupling criterion \eqref{decoupling criterion}. 
	We do this by evaluating the four-ellipsoid partition function of the parent theory using the localization formula of \cite{Hama:2012bg} and isolating contributions which we associate to string configurations. Exactly when the condition \eqref{decoupling criterion} is satisfied, we observe a  factorization of the contribution associated with the dynamics around the $K$-string to a product of two decoupled factors --  one is associated with the light modes in the bulk and the other describes the dynamics of the string moduli. We interpret this observed factorization as a sign for the decoupling of the string degrees of freedom from the bulk degrees of freedom. This provides an alternative derivation of the decoupling criterion \eqref{decoupling criterion}, which takes into account all the perturbative and non-perturbative quantum corrections.

	A brief review of the localization analysis, as well as an explanation of the precise sense in which the contributions factorize, can be found in section \ref{localization}. The detailed localization analysis will appear in an upcoming publication \citep{GGKK}, in which we use localization techniques to study the worldsheet theories of the strings that satisfy the decoupling criterion \eqref{decoupling criterion}.

	The bulk-string decoupling condition (\ref{decoupling criterion}) implies that for generic charges the minimal string is coupled to the bulk. However, at winding numbers such that (\ref{decoupling criterion}) is satisfied the $K$-string decouples from the bulk and an effective description in terms of a two-dimensional theory exists.  We argue that in this case the long-range modes are combined to create fewer modes with faster decay such that the string  decouples from the bulk. 
	
	Interestingly, the criterion (\ref{decoupling criterion}) is exactly the condition ensuring that there are no Aharanov-Bohm phases for particles circling the string -- exactly when (\ref{decoupling criterion}) is not satisfied, there exist a particle in the spectrum that acquires a non-trivial Aharonov-Bohm phase when circling the string.  Due to the bulk-string coupling, however, the  string and the particle cannot be treated as two separated objects,  and the Aharonov-Bohm experiment cannot be performed.

	The outline of the paper is as follows. In section \ref{u1twoflav} we demonstrate  the existence of long-range zero-modes and the  relation to the bulk-string decoupling  and Aharonov-Bohm phases in the simplest example of a $U(1)$ gauge theory with two flavors. We study the mixing of the bulk and string modes and argue that when long-range modes are absent the mixing terms are suppressed at low energies. In section \ref{sec:HT} we move on to non-abelian strings and review some of the well-known properties of the strings and their moduli spaces in the case where the gauge group is $U(N_c)$ and all the flavors have the same $U(1)$ charge. In section \ref{general gauging} we consider strings in the $SU(N_c)\times U(1)$ gauge theory with general $U(1)$ charges. We study the size zero-modes of these generalized strings and find that long-range modes exist if and only if \eqref{decoupling criterion} is not satisfied.  In section \ref{factorization} we discuss the mixing of the bulk and string modes and derive the decoupling criterion (\ref{decoupling criterion}) based on the size-modes analysis of section \ref{general gauging}.  We also show that the condition \eqref{decoupling criterion} coincides with the condition that all the Aharonov-Bohm phases are trivial.  In section \ref{Kstringmoduli} we comment on the size-modes of the  non-minimal string in cases where the minimal string is coupled to the bulk. In section \ref{localization} we sketch the non-perturbative derivation of the decoupling criterion \eqref{decoupling criterion}, which is based on a localization analysis that will appear in \cite{GGKK}.

\section{$U(1)$ with two Flavors}
\label{u1twoflav}

The simplest example of strings with size zero-modes arises in a $U(1)$ gauge theory with two charged flavors \citep{Vachaspati:1991dz, Hindmarsh:1992yy, Leese:1992fn,Gibbons:1992gt}. In this section, we will use this example to illustrate the mechanism behind the bulk-string factorization and the relation to the decay rate of the size-modes.  For our discussion, it will be important that the two flavors have different $U(1)$ charges. Other aspects of the string solution with general $U(1)$ charges were studied in \cite{Eto:2009bz}.  

The matter content consists of two charged scalars, denoted by $q_{1,2}$, with $U(1)$ charges $c_1=p>0$ and $c_2=1$. The Lagrangian of the theory is
\eq{\lag= \onov{4e^2}\left(F_{\mu\nu}\right)^2+\sum_i|\cD_\mu q_i|^2-\frac{e^2}{8}\left(\sum_i c_i|q_i|^2-\xi\right)^2-\mu^2|q_2|^2\ ,\ \;\;\;\cD_\mu q_i=(\partial_\mu-ic_iA_\mu)q_i\;,}
where $e$ is the gauge coupling and $\xi>0$. We will assume that $\mu\ll e\sqrt{\xi}$. The coefficient in front of the potential was chosen such that the theory admits a completion to an $\cN=2$ supersymmetric $U(1)$ gauge theory with two hypermultiplets and a Fayet-Iliopoulos parameter $\xi$. This fine-tuned coupling allows for a Bogmol'nyi completion, from which topologically stable strings can be constructed.\footnote{Let us stress that we have included only the minimal ingredients that are necessary to  demonstrate the idea of this section, and this is why we didn't include more fields or a mass term for $q_1$.  The more general case is a special case of the theory studied in section \ref{general gauging}. } 

The theory has a unique vacuum, which is given, up to gauge transformations, by
\eql{ourvacuum}{q_1=v\equiv \sqrt{\frac{\xi}{p}}\ ,\;\;\; q_2=0\ ,\;\;\; A_\mu=0\ .}  
In the vacuum \eqref{ourvacuum}, the gauge symmetry is broken down to $\mathbb{Z}_{p}$. The gauge field and the scalar $q_1$ obtain a mass $m_W=\frac{1}{\sqrt 2}pev$.

The string solutions can be constructed in the following way. We will choose the string to lie along the $x^3$ direction. The tension can be written as 
\eql{u1tension}{T=\int dx^1dx^2&\left[\onov{2e^2}\left(B_3\pm \frac{e^2}{2}\left(\sum_ic_i|q_i|^2-\xi\right)\right)^2+\sum_i|\cD_1q_i\pm i\cD_2q_i|^2+\mu^2|q_2|^2\pm  \xi B_3\right]\;,}
where $B_3=F_{12}$ and all other components of $F_{\mu\nu}$ were taken to be zero. We also assumed that the configuration depends only on the $x^1,\ x^2$ coordinates.
The $\pm$ sign labels the two possible choices for the sign of the magnetic flux.  
  
The first three terms in \eqref{u1tension} give the Bogomol'nyi equations for this string,
\eql{Bogmol'nyi-u(1)}{B_3\pm \frac{e^2}{2}\left(p|q_1|^2+|q_2|^2-\xi\right)=0\;,\;\;\; (\cD_1\pm i\cD_2)q_i=0\;,\;\;\;q_2=0\ ,}
while the last term is a topological term that measures the magnetic flux. 
We will focus on solutions with positive magnetic flux by taking the plus sign in \eqref{u1tension}. With this choice,
solutions to  (\ref{Bogmol'nyi-u(1)})  will satisfy
\eql{boundary-conditions-u(1)}{\begin{aligned}&\lim_{r\to\infty}q_1=ve^{ik\phi}\ ,\;\;\;k\in\mathbb{N}\;,\\&\bar{A}=-\frac{i}{p}\bar{\partial}\log q_1\ ,\\&q_2=0\ ,\end{aligned}}
where we have introduced holomorphic coordinates on the $x^1-x^2$ plane, 
\eq{z=x_1+ix_2\ ,\ \zb=x_1-ix_2\ ,\;\;\; \partial=\onov{2}(\partial_1-i\partial_2)\ ,\ \bar{\partial}=\onov{2}(\partial_1+i\partial_2)\ ,}
as well as polar coordinates, $z=re^{i\phi}$, $\bar z=re^{-i\phi}$.
Similarly, $\bar A\equiv \onov{2}(A_1+iA_2)$.

The total magnetic flux can be easily computed using Stokes' theorem, 
\eql{quantizedflux}{\Phi=\lim_{r\to\infty}\int_0^{2\pi} d\phi A_\phi=\frac{2\pi k}{p} \;.}
Indeed, due to the symmetry breaking pattern of the vacuum, the allowed fluxes are classified by $\pi_1\left(U(1)/\mathbb{Z}_p\right)$, which allows for fractional fluxes, quantized as in (\ref{quantizedflux}). 

Plugging equation (\ref{boundary-conditions-u(1)}) into the first equation in (\ref{Bogmol'nyi-u(1)}), one obtains a differential equation for $q_1$.  The boundary conditions (\ref{boundary-conditions-u(1)}) guarantee that  $q_1$ has $k$ zeros at positions $z_a$.  Close to the zeros it behaves as \eql{u1zeros}{\lim_{\vec{r}\to\vec{r}_a}q_1\sim z-z_a\ .}

The string solution described above has translational zero-modes, related to the locations of the zeros $z_a$. For positive $\mu$ these are the only exact zero-modes. However, we will be interested in fluctuations above the string solution with energies in the range $\mu\lesssim E\ll ev$. In this energy limit
 $A_\mu$ and $q_1$ decouple, while  $q_2$ is treated as a light dynamical field. 

The field $q_2$ has two types of excitations. The first are the vacuum excitations which are simply the excitations of $q_2$ above the vacuum \eqref{ourvacuum} and are expected to exist also far away from the string. The second are approximate string zero-modes, which become exact zero-modes in the limit $\mu\to 0$ where the Bogomol'nyi equations read 
\eql{Bogmol'nyi-u(1)mu0}{B_3+ \frac{e^2}{2}\left(p|q_1|^2+|q_2|^2-\xi\right)=(\cD_1+i\cD_2)q_i=0\ .}
These quasi-moduli exist if 
there exist non-trivial solutions to (\ref{Bogmol'nyi-u(1)mu0}) in which $q_2\neq 0$, which satisfy the boundary conditions (\ref{boundary-conditions-u(1)}).

The second equation in (\ref{Bogmol'nyi-u(1)mu0}) implies that 
\eq{\bar{A}=-\frac{i}{p}\bar{\partial}\log q_1\;,\;\;\; q_2=q_1^{1/p}f(z)\;,}
where $f(z)$ is independent of $\bar z$. Plugging this into the first equation in (\ref{Bogmol'nyi-u(1)mu0}) one finds a differential equation for $q_1$ that depends on the function $f(z)$. Let us now discuss the constraints that the boundary conditions (\ref{boundary-conditions-u(1)}) impose on  $f(z)$.  

For simplicity, let us assume that all the zeros of $q_1$ coincide at the origin such that $\lim\limits_{r\to 0}q_1\sim z^k$. Thus, close to the origin $q_2\sim z^{k/p}f(z)$. For $q_2$ to be well-defined at $r=0$, $z^{k/p}f(z)$ needs to have an expansion in non-negative powers of $z$. In addition, the boundary conditions (\ref{boundary-conditions-u(1)}) require that $\lim\limits_{r\to\infty} f(z)= 0$.  We thus conclude that \eql{q2zero}{q_2=q_1^{1/p}\sum_{n=0}^{\lceil k/p\rceil-1}\rho^{(n)}z^{n-k/p}\ ,}
where $\rho^{(n)}\in\mathbb{C}$ are arbitrary coefficients.\footnote{The generalization to the case where the zeros are distinct is straightforward and results in \eq{q_2=\left(\frac{q_1}{\prod_{a=1}^k(z-z_a)}\right)^{1/p}\sum_{n=0}^{\lceil k/p\rceil-1}\rho^{(n)}z^n\ .} See also section \ref{Kstring}, in which  this analysis is done for the more general $SU(N_c)\times U(1)$ gauge theories.} The quasi-moduli $\rho^{(n)}$ are related to the effective size of the string in the transverse plane. See appendix \ref{sizeofstring}. We will refer to these moduli as size-modes.
In the proceeding, we will study the effect of these zero-modes on the low-energy theory and show how the question of decoupling from the bulk is related to whether $k/p$ is  integer or not. 

In \eqref{q2zero}, the parameters $\rho^{(n)}$ are constants. However, to study the low-energy dynamics we should allow them to depend weakly on the worldsheet coordinates $x^I$, $I=0,3$. Let us split the low-energy $q_2$ excitations as  
\eql{q2expansion}{q_2=\chi(x^\mu)+q_1^{1/p}\sum_{n=0}^{\lceil k/p\rceil-1}\rho^{(n)}(x^I)z^{n-k/p}\ ,}
where $\chi(x^\mu)$ is the four-dimensional bulk-field and $\rho^{(n)}(x^I)$ are the string size-modes, which have been promoted to fields on the worldsheet. The distinction between the two types of excitations is meaningful only if the bulk and the string modes are decoupled. We will assume that this is true and show that the decoupling assumption leads to a contradiction if $\frac{k}{p}\notin\mathbb{Z}$.

Plugging the contribution of one size-mode,  $q_2\to q_1^{1/p}\frac{\rho(x^I)}{z^{\be}}$, into the $q_2$ kinetic term, and integrating over the transverse directions, we obtain a kinetic term on the worldsheet of the form\footnote{Note that the solution for $q_1$ depends on $\rho^{(n)}$ (see appendix \ref{sizeofstring} for more details) and therefore it also depends on the worldsheet coordinates $x^I$.}
\eql{kinint}{\int rdrd\phi r^{-2\be}\left|\partial_I  \left(q_1^{1/p} \rho\right)\right|^2\ ,\;\;\; I=0,3\ .}
For $\be\leq 1$ this integral is  IR divergent.  Thus, the slow-decaying size-modes are non-normalizable.  If these divergences are not regulated, the $\rho$ parameters cannot be promoted to dynamical worldsheet fields. For a discussion on how to deal with non-normalizable zero-modes see \cite{Intriligator:2013lca}. Moreover, due to the non-zero mass even constant non-vanishing values of $\rho$ are forbidden. This is because the contribution of the mass term to the tension is infinite due to a similar IR divergence. Recall, however, that we are not interested only in exact zero-modes, but also in  light modes which are dynamical at energies $\mu\lesssim E\ll ev$. Therefore, we can redefine the size-modes in the IR without ruining the fact that their contribution to the energy of the string vanishes at the $\mu\to 0$ limit. As in \citep{Shifman:2006kd}, we will treat the small mass $\mu$ as an effective IR regulator and consider $1/{\mu}$ as the upper limit of the integral over $r$.

Next, we will canonically normalize the fields on the worldsheet,\footnote{We took $q_1(\rho=0)$ in \eqref{canondef} to isolate the term which is quadratic in $\rho$. Terms coming from $q_1(\rho\neq 0)$ contribute to higher order terms in the effective low-energy worldsheet theory. }
\eql{canondef}{|\partial_I\rho'^{(n)}|^2=|\partial_I \rho^{(n)}|^2\int dr d\phi r^{1-2\be}|q_1|^{2/p}(\rho=0)\ .}
In terms of the canonically normalized fields, $\rho'^{(n)}$, equation \eqref{q2expansion} takes the form
\eql{q2expansion2}{q_2=\chi(x^\mu)+\left(\frac{q_1}{v}\right)^{1/p}\sum_{n=0}^{\lceil k/p\rceil-1}b_n\rho'^{(n)}(x^I)z^{n-k/p}\ ,}
where
\eql{317}{b_n=v^{1/p}\left(\int dr d\phi r^{1-2\be}|q_1|^{2/p}(\rho=0)\right)^{-\frac{1}{2}}\;.}

The string is decoupled from the bulk, and the expansion \eqref{q2expansion2} is a good expansion, if the terms that mix $\chi$ and $\rho'^{(n)}$ in the action are negligble at low energies. Consider, for example, the kinetic term 
\eql{mixedkinetic}{\int d^4x |\partial_I q_2|^2=&\int d^4 x|\partial_I\chi|^2+ \sum_n\int d^2x^I|\partial_I\rho'^{(n)}|^2\\&+\sum_n\left(\int d^4x \frac{b_n}{v^{1/p}z^{k/p-n}}\partial_I\chi^* \partial^I(q_1^{1/p}\rho'^{(n)})+\hbox{c.c.}\right)+\cO(\rho^3)\;.}
All the quadratic terms that mix two different string modes, $\rho'^{(n)},\rho'^{(m)}$ for $n\neq m$, vanish after integrating over $\phi$ due to their different angular dependence. 
Far away from the string core, we can approximate $q_1\sim ve^{ik\phi}$, and the mixing term obtains the form 
\eql{mixingterm}{e^{in\phi}\frac{b_n}{r^\be}\partial_I\chi^* \partial^I\rho'^{(n)}+\hbox{c.c.}\ ,\ \be=k/p-n\ .}

The coupling $b_n$ has mass dimension $1-\be_n$. Thus, the operator above is relevant at low energies if $\be_n<1$, irrelevant if $\be_n>1$, and classically marginal if $\be_n=1$.
Let us now estimate how the size of the coefficient $b_n$ scales with $v$ and $\mu$.
Consider the integral \eql{bintegral}{\int_0^{\frac{1}{\mu}} dr r^{1-2\be}|q_1|^{2/p}(\rho=0)\ .} We are interested in the parametric dependence of this integral on the two scales $\mu$ and $v$, where we are only interested in the leading term in the $\frac{\mu}{v}\to 0$ limit.\footnote{We don't distinguish here between $v$ and $m_W$ which are both taken to be very large compared to $\mu$. } Note that the $\mu$ dependence of (\ref{bintegral}) enters only through the upper limit of the  integral. When $r$ is large enough, much larger than the transverse size of the string, we can approximate $|q_1|=v$. Using this fact, the separation of scales $\mu\ll v$ and dimensional analysis, one can show that the leading dependence behaves as $v^{2/p}\mu^{2\be-2}$ for $\be<1$, $v^{2/p}\log(v/\mu)$ for $\be=1$ and  $v^{2\be-2+2/p}$ for $\be>1$. We thus find
\eql{bresults}{b_{\be>1}\sim v^{1-\be}\ ,\;\;\; b_{\be=1}\sim \onov{\sqrt{\log(v/\mu)}}\ ,\;\;\; b_{\be<1}\sim \mu^{1-\be}\ .}

This leads us to conclude that at low energies, $E\ll v$, the $\be>1$ modes decouple from the bulk modes. For $\be=1$, the classically marginal  interaction term is suppressed in the $\frac{\mu}{v}\to 0$ limit.  Thus, if all the modes decay at least as fast as $\frac{1}{r}$, the string and the bulk decouple at low energies, and an effective description in terms of a two-dimensional theory is appropriate.  
On the other hand, modes with $\beta<1$ are coupled to the bulk even at low energies.\footnote{By low energies we mean $\mu<E\ll v$. At energies below $\mu$ the string  moduli 
	cannot be excited and the bulk is empty.} Thus, the existence of such modes is in contradiction with the decoupling assumption. As can be seen from \eqref{q2expansion2}, these modes exist iff $\frac{k}{p}\slashed{\in}\mathbb{Z}$;  when $\frac{k}{p}
\in\mathbb{Z}$ all the modes decay at least as fast as $\onov{r}$. To conclude, only when $\frac{k}{p}\in\mathbb{Z}$, the string and the bulk decouple at low energies.

So far we discussed only the kinetic term $|\partial_I q_2|^2$. Using the fact that $\rho'^{(n)}$ always appear with a $b_n$ factor, similar arguments show that the conclusions hold also when the other terms are included.
While this analysis is purely classical, the conclusions are supported by a non-perturbative localization computation which will be reviewed in section \ref{localization}.

When  evaluating the integral \eqref{bintegral}, we  assumed that $\mu r_s\ll1$ where $r_s$ is the transverse size of the string as defined in equation \eqref{rstorho} of appendix \ref{sizeofstring}. This assumption is necessary if we want to approximate $|q_1|=v$ at large distances $r\sim\onov{\mu}$. However, $r_s$ is a dynamical parameter and we need to verify that this assumption is consistent. Using \eqref{rstorho} and \eqref{canondef}, we can express $r_s$ as
\eql{rsbesmall}{r_s\sim\begin{cases}  \onov{\mu}\left(\frac{\mu}{v}\right)^{1/\be_{\text{min}}}|\rho'_{\text{min}}|^{1/\be_{\text{min}}}&\quad \text{if } \be_{\text{min}}<1\;,\\
		\onov{v}\frac{|\rho'_{\text{min}}|}{\sqrt{\log(v/\mu)}}&\quad \text{if } \be_{\text{min}}=1\;. \end{cases}}
Thus, $\mu r_s\ll 1$ if $|\rho'_{\text{min}}|\ll v/\mu$. This is indeed satisfied as we are interested in fluctuations with energy $E\ll v$. Therefore, this assumption is consistent at low energies.

We also want to make the following observation. When a $q_2$-particle  circles the string, it acquires an Aharonov-Bohm phase 
\eq{\lim_{r\to\infty}\int d\phi A_\phi=\frac{2\pi k}{p}\ .}
This phase is trivial iff $k/p\in\mathbb{Z}$. 

To conclude, one finds the following correlation between long-range modes, bulk-string decoupling and Aharonov-Bohm phases.
When $k/p\in\mathbb{Z}$,
\begin{enumerate}
	\item The slowest-decaying zero-mode decays like $\onov{r}$.
	\item The string and the bulk are decoupled at low energies.
	\item $q_2$ does not acquire an Aharonov-Bohm phase when circling the string.
\end{enumerate}
On the other hand, when $k/p\ \slashed{\in}\mathbb{Z}$,
\begin{enumerate}
	\item The slowest-decaying zero-mode decays like $\onov{r^\be}$ with $0<\be<1$.
	\item The string and the bulk are coupled even at low energies.
	\item $q_2$ acquires a non-trivial Aharonov-Bohm phase when circling the string.
\end{enumerate}

In the next sections, we will generalize this analysis to non-abelian strings in $SU(N_c)\times U(1)$ gauge theory. We will find that long-range size-modes exist if \eqref{decoupling criterion} is not satisfied. We will then use the analysis we presented in this section to argue that the bulk and the string decouple if the long-range modes are absent, that is, if \eqref{decoupling criterion} is satisfied.   We will also find that the same correlation between Aharonov-Bohm phases and long-range zero-modes holds in the more general case.

\section{Non-Abelian Vortex-Strings -- Review of the Literature (equal $U(1)$ charges)}
\label{sec:HT}

We now move on to non-abelian strings. We will start by reviewing the well-studied case of $U(N_c)$ gauge group, where all the flavors have the same charge under the $U(1)$ factor of the gauge group.  In the next section we will generalize the set-up by allowing different flavors to have different $U(1)$ charges.   This section is based on many references.
See, for example, \citep{Auzzi:2003fs,Hanany:2003hp,Hanany:2004ea,Shifman:2004dr,Shifman:2006kd}, and the reviews \cite{Shifman:2007ce,Tong:2005un,Tong:2008qd}.

Our starting point  is a four-dimensional $\mathcal{N}=2$ supersymmetric theory with $U(N_c)\equiv SU(N_c)\times U(1)/\mathbb{Z}_{N_c}$ gauge symmetry and $N_f\geq N_c$ hypermultiplets in the fundamental representation of the gauge group.\footnote{If $N_f<N_c$ the theory doesn't have a  supersymmetric vacuum in the presence of a Fayet-Iliopoulos  term.}   The theory is parametrized by the complexified gauge couplings,
\eql{tau}{\tau_{u(1)}=\frac{\te_{u(1)}}{2\pi}+\frac{4\pi i}{e^2}\;,\;\;\;  \tau_{su(N)}=\frac{\te}{2\pi}+\frac{4\pi i}{g^2}\;,}
the hypermultiplet masses $\mu_i$, $i=1,...,N_f$, and a Fayet-Iliopoulos (FI) parameter $\xi>0$. We assume that the masses are non-degenerate ($\mu_i\neq\mu_j$ if $i\neq j$) and that the mass differences are small, $|\mu_i-\mu_j|\ll e\sqrt{\xi}, g\sqrt{\xi}$ for all $i,j$.

The bosonic action for this theory is given by 
\eql{N2action}{S&=\int d^4x\left[ \onov{4g^2}\left(F_{\mu\nu}^\al\right)^2+\onov{4e^2}\left(F_{\mu\nu}'\right)^2+\onov{g^2}|\cD_\mu a^\al|^2+\onov{e^2}|\partial_\mu a'|^2+\sum_i|\cD_\mu q_i|^2+\sum_i|\cD_\mu \tilde{q}^i|^2-V +\lag_{\te}   \right]\ , \\
	\lag_{\te}&=\frac{\te}{32\pi^2}F_{\mu\nu}^\al\tilde{F}^{\al\mu\nu}+\frac{\te_{u(1)}}{32\pi^2}F'_{\mu\nu}\tilde{F}'^{\mu\nu} \ , \\
	V&=\frac{g^2}{2}\left(\onov{g^2}f^{\al\be\gamma}a_\be^{\dagger}a_\gamma+\sum_iq^{i\ \dagger}\la^\al q_i-\sum_i\tilde{q}^i\la^\al \tilde{q}_i^{\ \dagger}\right)^2+\frac{e^2}{8}\left(\sum_iq^{i\dagger}q_i-\sum_i\tilde{q}_i\tilde{q}^{i\dagger}-N_c\xi\right)^2\\
	&+2g^2\Big|\sum_i\tilde{q}^i\la^\al q_i\Big|^2+\frac{e^2}{2}\Big|\sum_i\tilde{q}^iq_i\Big|^2+\sum_i\Big|(a'+\la^\al a^\al-\mu_i)q_i\Big|^2+\sum_i\Big|(a'+\la^\al a^\al-\mu_i)\tilde{q}_i^\dagger\Big|^2\;,}
where
\eq{ \cD_\mu q_i&=(\partial_\mu-iA_\mu'-i\la^\al A_\mu^\al)q_i,\;\;\; \cD_\mu \tilde{q}^\dagger_i=(\partial_\mu-iA_\mu'-i\la^\al A_\mu^\al)\tilde{q}^\dagger_i,\;\;\; \cD_\mu a^\al=(\partial_\mu\delta^{\al\gamma}-if^{\al\be\gamma}A_\be)a_\gamma\ .}
Here $a',\ A_\mu'$ are the scalar and gauge field of the $U(1)$ gauge multiplet, and $a^\al,\ A_\mu^\al$, $\al=1,...,N_c^2-1$, are the scalars and gauge fields of the $SU(N_c)$ gauge multiplet. $\la^\al$ and $f^{\al\be\gamma}$ are the $SU(N_c)$ generators and structure constants respectively. The hypermultiplets scalars are denoted by $q_i^a\ ,\ \tilde{q}^{i}_a$ where $i=1,..., N_f$ is the flavor index. The color index 
,  $a=1,...,N_c$, was suppressed in (ֿ\ref{N2action}). $q$ transforms in the fundamental representation of $U(N_c)$, while $\tilde q$ transforms in the antifundamental representation. Below,   we will also use a matrix notation, \eq{\textbf{q}=\mat{q_1^1&q_2^1&\cdots&q_{N_f}^1\\q_1^2&q_2^2&\cdots&q_{N_f}^2\\\vdots&\vdots&\ddots&\vdots\\q_1^{N_c}&q_2^{N_c}&\cdots&q_{N_f}^{N_c}}\ ,\ \tilde{\textbf{q}}=\mat{\tilde{q}_1^1&\tilde{q}_2^1&\cdots&\tilde{q}_{N_c}^1\\\tilde{q}_1^2&\tilde{q}_2^2&\cdots&\tilde{q}_{N_c}^2\\\vdots&\vdots&\ddots&\vdots\\\tilde{q}_1^{N_f}&\tilde{q}_2^{N_f}&\cdots&\tilde{q}_{N_c}^{N_f}}\ .}

In the absence of mass and FI terms the theory has a $PSU(N_f)$  global symmetry and an $SU(2)_R\times U(1)_R/\mathbb{Z}_2$ $R$-symmetry. The  $U(1)_R$ symmetry is anomalous. In the presence of generic masses the global symmetry is broken explicitly to $U(1)^{N_f-1}$. However, we are interested in the $|\mu_i-\mu_j|\ll g\sqrt{\xi},\ e\sqrt{\xi}$ limit, in which we can treat the global $PSU(N_f)$ symmetry of the massless theory as an approximate symmetry. 
The FI term explicitly breaks the $SU(2)_R$ symmetry to its Cartan but does not break the $\cN=2$ supersymmetry. We will denote the $U(1)\subset SU(2)_R$ that is preserved in the presence of the FI term by $U(1)_J$.

The FI term forces the vacua of the theory to sit in the Higgs branch. The theory has ${N_f}\choose{N_c}$ vacua, labeled by choices of $N_c$ out of the $N_f$ hypermultiplets. The vacuum labeled by $1,2,..., N_c$ is given, up to gauge transformations,  by
\eql{vacuum}{\begin{aligned}
		&\textbf{q}=\mat{\sqrt{\xi}\,\mathbb{I}_{N_c\times N_c}&0_{(N_f-N_c)\times N_c}}\;,\\
		&\delta_{ab}a'+\la^\al_{ab}a^\al=\delta_{ab}\mu_b\;,\;\;\;a,b=1,...,N_c\;,
	\end{aligned}}
	and all other fields vanish.
	This vacuum is invariant under the transformations
	\eql{vac_symm}{\textbf{q}\rightarrow U^\dagger \textbf{q}\mat{U&0\\0&V}\ ,\ U\in U(N_c)\ ,\ V\in U(N_f-N_c)\ ,\ \det U\cdot \det V=1\ .}
	Thus, the gauge symmetry is broken completely, and the residual (approximate) global symmetry is
	$S[U(N_c)\times U(N_f-N_c)]$.
	The vacuum also preserves a $U(1)$ $R$-symmetry, in which the $U(1)_J$ transformation of $q$ is canceled by a $U(1)$ gauge transformation.

	The spectrum in the vacuum consists of $2N_c(N_f-N_c)$ pseudo-Goldstone-bosons,   which reside in the $N_c(N_f-N_c)$ light hypermultiplets labeled by  ${i>N_c}$.   In addition, the $N_c^2$ gauge multiplets combine with the $N_c^2$ hypermultiplets labeled by ${i\leq N_c}$ via the Higgs mechanism to create $N_c^2$ long massive vector multiplets, with masses $m_{W}\propto e\sqrt{\xi}, g\sqrt{\xi}$.
	
	As in the previous section, we derive the equations for a string lying along the $x^3$ direction using a Bogomol'nyi completion of the tension. Assuming that the configuration does not depend on the worldsheet coordinates $x^0$ and $x^3$, and that $\tilde q$, $A_0$, $A_3$ and the off-diagonal elements of $a$ vanish, the tension can be written as  
	\eql{tension}{T=\int dx^1dx^2&\left[\onov{2e^2}\left(B'_3\pm \frac{e^2}{2}\left(\sum_i|q_i|^2-N_c\xi\right)\right)^2+\onov{2g^2}\left(B_3^\al\pm g^2\sum_iq_i^\dagger\la^\al q_i\right)^2+\sum_i|\cD_1q_i\pm i\cD_2q_i|^2\right.\\&\left.\;+\onov{g^2}|\cD_\mu a^\al|^2+\onov{e^2}|\partial_\mu a'|^2+\sum_i|(a'+\la^\al a^\al-\mu_i)q_i|^2\pm  N_c\xi B'_3\right]\\\geq \pm\int dx^1d&x^2 N_c\xi  B'_3}   
	with $B_3'=F_{12}'\ ,\ B_3^\al=F_{12}^\al$.
	The last term is a topological term that measures the magnetic flux.  Configurations for which all the other terms vanish minimize the tension in a given topological sector and give rise to $\frac{1}{2}$-BPS strings. 
	
	Let us now focus on string solutions with one quanta of positive magnetic flux. To find the solution, one uses the ansatz \eql{basicv}{\textbf{q}=\mat{q'(r)&0&0&\cdots&0&0&\cdots&0\\0&q'(r)&0&\cdots&0&0&\cdots&0\\0&0&q'(r)&\cdots&0&0&\cdots&
			0\\\vdots&\vdots&\vdots&\ddots&\vdots&\vdots&\ddots&\vdots\\0&0&0&\cdots&q_v(r) e^{i\phi}&0&\cdots&0}\ ,}
	where the profile functions need to satisfy the boundary conditions \eql{boundaryconditions}{\lim_{r\to\infty}q'(r)=\lim_{r\to\infty}q_v(r)=\sqrt{\xi}\ .} The boundary conditions together with the equation $(\cD_1+i\cD_2)q=0$ imply that $q'(r)$ has no zeros, while $q_v(r)$ has a single zero at $r=0$. Close to the zero it behaves as $q_v(r)\sim r$ \cite{Taubes:1979tm}.\footnote{The behavior near $r=0$ can be easily understood in the following way. At the origin $q_v(r)$ must go to zero  because of the angular dependence, so we can write $q_v(r)\sim r^{\Delta}$ near the origin. From equation \eqref{flux} the $U(1)$ gauge field near the origin is $A_\phi=1-r\dr \log(r^{\Delta})+\cO(r)$. In a non-singular string solution the flux through a circle of zero radius must vanish. Thus, $\Delta=1$.}  The gauge multiplet scalars do not play any role in the solution; they are fixed on their vacuum expectation value given in \eqref{vacuum}.  This guarantees the vanishing of the terms in \eqref{tension} that involve  $a$ and $a'$. We are thus   left with the Bogmol'nyi equations:
	\begin{align}
	&B'_3+ \frac{e^2}{2}\left(\sum_i|q_i|^2-N_c\xi\right)=0\;,\label{Bog1}\\&B_3^\al+ g^2\sum_iq_i^\dagger\la^\al q_i=0\;,\label{Bog2}\\&\cD_1q_i+ i\cD_2q_i=0\;.\label{Bog3}
	\end{align} 
	
	The Cartan gauge fields need to be turned on, with a profile which is determined from the functions $q'(r)$ and $q_v(r)$ using \eqref{Bog3}. Plugging this into (\ref{Bog1}-\ref{Bog2}) one obtains a set of differential equations for the profile functions $q'(r)$ and $q_v(r)$,  for which there is no known analytic solution. The solution has been studied numerically in \cite{Auzzi:2003fs} for $N_c=2$.

	Using equation \eqref{Bog3}  we can express the $U(1)$ gauge field as
	\eql{flux}{N_cA'_\phi=1-r\dr\log(q'(r)^{N_c-1}q_v(r))\ .}
	At $r\rightarrow\infty$, the second term goes to zero and the total $U(1)$ magnetic flux and tension are \eq{\Phi=\lim_{r\to\infty}\int d\phi A'_\phi=\frac{2\pi}{N_c}\ ,
		\;\;\;\ T=N_c\xi\Phi=2\pi\xi\ .} This flux is supported by the first homotopy group $\pi_1\left(U(1)\times SU(N_c)/\mathbb{Z}_{N_c}\right)$. 
	Similarly, the asymptotic values of the $SU(N_c)$ Cartan gauge fields can be computed. One finds that the Aharonov-Bohm phases for particles circling this string are trivial, that is, integer multiples of $2\pi$ (see section \ref{factorization}).

	Let us now discuss the zero-modes of the string solution. 
	The exact zero-modes are two center-of-mass zero-modes related to the two translations that are broken by the string solution. As in the previous section, we will also be interested in modes that are light with respect to $e\sqrt{\xi}$, $g\sqrt{\xi}$.  Therefore, we will study the zero-modes in the massless case. These will be treated as approximate zero-modes at the $|\mu_i-\mu_j|\ll g\sqrt{\xi},\ e\sqrt{\xi}$ limit.
	
	Acting with the $S[U(N_c)\times U(N_f-N_c)]$ symmetry of the vacuum,ֿ\footnote{Due to the winding of the string solution, the asymptotic symmetry in the string background does not act exactly as  \eqref{vac_symm}. Instead, the transformation \eqref{vac_symm} acts on  $\hat{\textbf{q}}=\text{diag}(1,1,...,1,e^{-i\phi})\textbf{q}$.} and dividing by the stabilizer we obtain $2(N_c-1)$ zero-modes which span the moduli space \eql{orient}{\frac{S[U(N_c)\times U(N_f-N_c)]}{S[U(N_c-1)\times U(1)\times U(N_f-N_c)]}=\frac{U(N_c)}{U(N_c-1)\times U(1)}=\mathbb{CP}^{N_c-1}\;.}  These will be referred  to as the orientation zero-modes.  
	
	When $N_f>N_c$, there exist extra approximate zero-modes,  in which the extra flavors play a role in the solution. In the presence of non-degenerate masses such modes cannot be exact zero-modes; a solution of the form \eqref{basicv} fixes the gauge multiplet scalars completely (as in \eqref{vacuum}) leaving us with mass terms of the form  $(\mu_i-\mu_a)^2|q_{i}^a|^2$ for the non-diagonal components of $\mathbf{q}$.  However, since the mass differences are small compared to $g\sqrt{\xi}$, $e\sqrt{\xi}$,  we will ignore these terms and allow any solution of the Bogomol'nyi equations (\ref{Bog1}-\ref{Bog3}), as long as the solution satisfies the same boundary conditions as the original string solution. 
	
	Assuming an ansatz of the form (\ref{basicv}) for $i\leq N_c$, but allowing the entries for $i>N_c$ to be non-zero,  equation  \eqref{Bog3} implies that \eq{q_{i>N_c}^{a<N_c}=q'(r)g_i^a(z)\ ,\;\;\; q_{i>N_c}^{N_c}=q_v(r)e^{i\phi}f_i(z)\;,}
	where $g_i^a(z)$, $f_i(z)$ are independent of $\bar z$. As in the previous section,  these functions are constrained by the boundary conditions, which  imply that $g_i^a(z)=0$ and $f_i(z)\propto\frac{1}{z}$.\footnote{The boundary conditions \eqref{boundaryconditions} and $\lim\limits_{r\to\infty} {q_{i>N_c}^a}=0$ imply that $\lim\limits_{r\to\infty}f_i(z)=\lim\limits_{r\to\infty}g_i^a(z)=0$. However, a holomorphic function that decays at infinity needs to have a singularity in the bulk. $q'(r)$ has no zeros and thus we conclude that $g_i^a(z)=0$. $q_v(r)$ has one zero at the origin. From the behavior near the origin we conclude, as in the discussion before equation \eqref{q2zero}, that  $f_i(z)\propto\frac{1}{z}$.  } We thus get,
	\eq{q_{i>N_c}^{a<N_c}=0\;,\;\;\;\;q_{i>N_c}^{N_c}=\frac{q_v(r)}{r}\rho_i\; ,\;\;\; \rho_i\in\mathbb{C}\;.}
	Plugging this into equations (\ref{Bog1}-\ref{Bog2}), we obtain differential equations for the profile functions $q'(r)$ and $q_v(r)$, which depend on $\sum_{i=N_c+1}^{N_f}|\rho_i|^2$.   The modulus $\sum_{i=N_c+1}^{N_f}|\rho_i|^2$ determines the effective size of the string in the radial direction. See, for example, \cite{Shifman:2006kd}  or appendix \ref{sizeofstring}. The $N_f-N_c$ complex moduli $\rho_i$ are thus referred  to as the size-modes.
	
	Naively, one would expect the size-modes to interact with the light bulk fields -- the pseudo-Goldstone-bosons and their superpartners. However, as shown in the previous section, modes that decay as ${1}/{r}$ decouple from the bulk at low-energies.\footnote{Note that, as in the previous section, one first has to regularize these modes by modifying the solution at distances above $\sim\frac{1}{|\mu_i-\mu_a|}$.} See section \ref{factorization} for more details. Thus, the bulk and the string modes decouple at energies $\ll e\sqrt{\xi}, g\sqrt{\xi}$. As a result, the low-energy dynamics in the string background is captured by a two-dimensional theory on the string worldsheet, in addition to decoupled bulk fields. 
	
 The two-dimensional theory on the worldsheet has $\cN=(2,2)$ supersymmetry, inherited from the residual supersymmetry in the string background. In addition, the four-dimensional  $U(1)$  $R$-symmetry in the string background and  the rotation symmetry in the plane transverse to the string both give rise to conserved vectorlike $R$-symmetries on the worldsheet. Thus, the worldsheet theory preserves a continuum of $R$-symmetries.
	
	 Based on various arguments, such as brane construction \cite{Hanany:2004ea,Hanany:2003hp}, matching of the BPS spectra \cite{Dorey:1998yh,Dorey:1999zk}, as well as explicit derivations in field theory \cite{Shifman:2004dr,Shifman:2006kd,Auzzi:2003fs,Hanany:2004ea},  the worldsheet theory has been identified as the low-energy limit of a GLSM with $U(1)$ gauge group. The matter content of this theory is one neutral chiral multiplet, parametrizing the center-of-mass moduli and their superpartners, $N_c$ positively charged chiral multiplets, parametrizing the orientation moduli and their superpartners, and $N_f-N_c$ negatively charged chiral multiplets, parametrizing the size-moduli $\rho_i$ and their superpartners. The complexified FI parameter of the two-dimensional theory is given in terms of the four-dimensional complexified $SU(N_c)$ gauge coupling, by $\frac{\te_{2d}}{2\pi }+i\xi_{2d}=\tau_{su(N_c)}$. The charged chiral multiplets inherit small masses $\sim \mu_i\ll e\sqrt{\xi}, g\sqrt{\xi}$, reflecting the fact that the orientation and size-modes are only approximate moduli. 
	
	While strings with winding number $K>1$ are harder to study, the brane construction of \cite{Hanany:2003hp} offers a generalization of the worldsheet GLSM to  a $U(K)$ gauge theory with one adjoint chiral multiplet, $N_c$ fundamental chiral multiplets and $N_f-N_c$ antifundamental chiral multiplets. The string solution and the moduli space have also been studied in field theory. For example, see \cite{Auzzi:2005gr,Eto:2005yh,Hashimoto:2005hi,Eto:2006cx}.

\section{General $U(1)$ Charges}\label{general gauging}
In this section we generalize the set-up of the previous section by allowing different $U(1)$ charges for the different hypermultiplets. Thus, we start from  a four-dimensional $\cN=2$ supersymmetric theory with $SU(N_c)\times U(1)$ gauge symmetry and  $N_f\geq N_c$ hypermultiplets in the fundamental representation of $SU(N_c)$. The $U(1)$ charges of the hypermultiplets will be denoted by $c_i\in\mathbb{Z}$, $i=1,...,N_f$. As in the previous section, we will introduce an FI parameter $\xi$ and non-degenerate hypermultiplet masses $\mu_i\ll e\sqrt{\xi},g\sqrt{\xi}$,\footnote{The mass $\mu_i$ in itself is unphysical if $c_i\neq 0$, as its value can be shifted arbitrarily by redefinitions of $a'$. What we actually need to assume is that the physical combinations that appear in equation \eqref{mass-in-vac}  are small with respect to $e\sqrt{\xi},g\sqrt{\xi}$.} where $e$ and $g$ are, respectively, the gauge couplings for the $U(1)$ and $SU(N_c)$ factors of the gauge group.

The bosonic part of the action is given by
\eql{generalaction}{S&=\int d^4x\left[ \onov{4g^2}\left(F_{\mu\nu}^\al\right)^2+\onov{4e^2}\left(F_{\mu\nu}'\right)^2+\onov{g^2}|\cD_\mu a^\al|^2+\onov{e^2}|\partial_\mu a'|^2+\sum_i|\cD_\mu q_i|^2+\sum_i|\cD_\mu \tilde{q}^i|^2-V+\lag_{\te}    \right]\ , \\
	\lag_{\te}&=\frac{\te}{32\pi^2}F_{\mu\nu}^\al\tilde{F}^{\al\mu\nu}+\frac{\te_{u(1)}}{32\pi^2}F'_{\mu\nu}\tilde{F}'^{\mu\nu} \ , \\
	V&=\frac{g^2}{2}\left(\onov{g^2}f^{\al\be\gamma}a_\be^{\dagger}a_\gamma+\sum_iq^{i\ \dagger}\la^\al q_i-\sum_i\tilde{q}^i\la^\al \tilde{q}_i^{\ \dagger}\right)^2+\frac{e^2}{8}\left(\sum_ic_iq^{i\dagger}q_i-\sum_ic_i\tilde{q}_i\tilde{q}^{i\dagger}-N_c\xi\right)^2\\
	&+2g^2\Big|\sum_i\tilde{q}^i\la^\al q_i\Big|^2+\frac{e^2}{2}\Big|\sum_ic_i\tilde{q}^iq_i\Big|^2+\sum_i\Big|(c_ia'+\la^\al a^\al-\mu_i)q_i\Big|^2+\sum_i\Big|(c_ia'+\la^\al a^\al-\mu_i)\tilde{q}_i^\dagger\Big|^2\;,}
where
\eq{ \cD_\mu q_i&=(\partial_\mu-ic_iA_\mu'-i\la^\al A_\mu^\al)q_i\ ,\ \cD_\mu \tilde{q}^\dagger_i=(\partial_\mu-ic_iA_\mu'-i\la^\al A_\mu^\al)\tilde{q}^\dagger_i\ ,\ \cD_\mu a^\al=(\partial_\mu\delta^{\al\gamma}-if^{\al\be\gamma}A_\be)a_\gamma\ .}
The notations are explained right after equation \eqref{N2action}. 

The theory enjoys a non-anomalous $U(1)_J\subset SU(2)_R$ $R$-symmetry and a $U(1)^{N_f-1}$ global symmetry. If some of the $U(1)$ charges $c_i$ are  equal, the global symmetry is enlarged in the massless limit to a non-abelian symmetry. 

When not all the $U(1)$ charges are the same ($c_i\neq c_j$ for some $i\neq j$), two classes of supersymmetric vacua exist: mesonic and baryonic. This is because the vacuum equations can be satisfied by giving a vacuum expectation value (VEV) to some $SU(N_c)$ invariant operator which is charged under the $U(1)$ gauge symmetry. When both charged mesons and charged baryons exist, there are different vacua corresponding to a meson getting a VEV or to a baryon getting a VEV.

The mesonic vacuum in which $q_i\tilde{q}^j$ obtains a VEV is given, up to gauge transformations, by
\eq{q_i^a=\tilde{q}_a^j=\sqrt{\frac{N_c\xi}{c_i-c_j}}\; ,}
for some choice of $a,i,j$ such that  $c_i>c_j$ if $\xi>0$ and $c_i<c_j$ if $\xi<0$.
In addition, $a'$ and one component of the non-abelian adjoint scalar $a$ should be turned on to satisfy
\eql{mes-vac-a}{c_ia'+\la_{aa}^\al a^\al-\mu_i=c_ja'+\la_{aa}^\al a^\al-\mu_j=0\ .}
Note that the index $a$ is not summed over in \eqref{mes-vac-a}. All the other fields vanish in the mesonic vacuum. 
The gauge symmetry is broken in this vacuum to $\mathbb{Z}_{|c_i-c_j|}\times SU(N_c-1)$. The study of the mesonic vacua and the corresponding strings is left for a 
future work.

In this paper we will study the baryonic vacua and  the corresponding strings. 
To construct the baryonic vacuum, which generalizes the  vacuum \eqref{vacuum}, one needs to choose $N_c$ out of the $N_f$ hypermultiplets such that the sum of their charges is not zero. In general, different choices of $N_c$ hypermultiplets lead to different types of vacua and different types of strings. Without loss of generality, we will choose the first $N_c$ hypermultiplets and assume that $C\equiv\sum_{i=1}^{N_c}c_i>0$ and $\xi>0$. The case of $C,\xi<0$ is exactly the same, while the cases of $C<0<\xi$ and $\xi<0<C$ become the same after the replacement $q\leftrightarrow\tilde{q}$.

The vacuum for this choice of hypermultiplets is given, up to gauge transformations, by
\eql{generalvac}{\begin{aligned}
		&\textbf{q}=\mat{v\mathbb{I}_{N_c\times N_c}&0_{(N_f-N_c)\times N_c}}\; ,\;\;\;v^2=\frac{N_c\xi}{C}\; \;,\\  &c_b\delta_{ab}a'+\la^\al_{ab}a^\al=\mu_b\delta_{ab}\;, \;\;\;\;\;\;\;\;  a,b=1,...,N_c\;.
	\end{aligned}}
	All the other fields vanish in the vacuum.
	The masses of the $q_{i}^a$ excitations in the vacuum (\ref{generalvac}) are shifted by the vacuum expectation values of the vector multiplet scalars, and are given by
	\eql{mass-in-vac}{\mu_i-\mu_a+\frac{c_a-c_i}{C}\sum_{b=1}^{N_c}{\mu_b}\;.}

	The vacuum  \eqref{generalvac} is invariant under the gauge transformation
	\eq{q_i^a\rightarrow e^{2\pi im \frac{c_i-c_a}{C}}q_i^a\;,\;\;\;m=1,2,...,C\;.}
	Thus, the gauge symmetry is broken in the vacuum  to $\mathbb{Z}_C$. 
	The vacuum preserves a $U(1)$ $R$-symmetry, under which $q_{i}^a$ has charge $N_c\frac{c_a-c_i}{C}$ (this symmetry is a combination of the $U(1)_J$ $R$-symmetry and an $SU(N_c)\times U(1)$ transformation).  Due to the residual discrete $\mathbb{Z}_C$ gauge symmetry, the relevant homotopy group supporting strings above this vacuum is $\pi_1(U(1)\times SU(N_c)/\mathbb{Z}_C)$ which allows strings with fluxes quantized as \eq{\Phi=\lim_{r\to\infty}\int d\phi A'_\phi=\frac{2\pi m}{C}\;, \;\;\; m\in\mathbb{Z}.}
	As before, the spectrum in the vacuum consists of $N_c^2$ long massive vector multiplets and $N_c(N_f-N_c)$ light hypermultiplets, with masses given by equation \eqref{mass-in-vac} for $i=N_c+1,...,N_f$, $a=1,...,N_c$. 
	
	Note that the global structure of the gauge group has no effect on the spectrum of strings.  For example,  in the previous section we could have taken   the gauge group to be $SU(N_c)\times U(1)$ instead of  $U(N_c)={SU(N_c)\times U(1)}/{\mathbb{Z}_{N_c}}$.  More generally, if $\frac{c_i-c_a}{C}\in\mathbb{Z}$ for all $i,a$, one can take the gauge group to be $SU(N_c)\times U(1)$ or ${SU(N_c)\times U(1)}/{\mathbb{Z}_{C}}$. The vacuum \eqref{generalvac} preserves    a $\mathbb{Z}_C$ subgroup in the first case and breaks the gauge symmetry completely in the second case.  The relevant homotopy group  is $\pi_1\left({SU(N_c)\times U(1)}/{\mathbb{Z}_{C}}\right)$ in both cases and the spectrum of strings is the same.

	To construct the Bogomol'nyi equations,
	we repeat the procedure of the previous section with the modification of different charges for the different hypermultiplets. We will keep the $\tilde{q}$ fields here. We will see that, depending on the charges $c_i$, these fields may play a role in the string moduli space.
	The analogue of \eqref{tension} is 
	\eql{tension-general}{T=\int dx^1dx^2&\left[\onov{2e^2}\left(B'_3\pm \frac{e^2}{2}\left(\sum_ic_i|q_i|^2-\sum_ic_i|\tilde{q}_i|^2-N_c\xi\right)\right)^2\right.\\&\left.+\onov{2g^2}\left(B_3^\al\pm g^2\sum_iq^{i\dagger}\la^\al q_i\mp g^2\sum_i\tilde{q}^i\la^\al\tilde{q}_i^\dagger\right)^2+\sum_i\Big|\cD_1q_i\pm i\cD_2q_i\Big|^2+\sum_i\Big|\cD_1\tilde{q}_i\pm i\cD_2\tilde{q}^i\Big|^2\right.\\&\left.+2g^2\Big|\sum_i\tilde{q}^i\la^\al q_i\Big|^2+\frac{e^2}{2}\Big|\sum_ic_i\tilde{q}^i q_i\Big|^2+\onov{g^2}\Big|\cD_\mu a^\al\Big|^2+\onov{e^2}\Big|\partial_\mu a'\Big|^2\right.\\&\left.+\sum_i\Big|(c_ia'+\la^\al a^\al-\mu_i)q_i\Big|^2+\sum_i\Big|(c_ia'+\la^\al a^\al-\mu_i)\tilde{q}_i^\dagger\Big|^2ֿ\pm  N_c\xi B'_3\right]\\&\geq \pm\int dx^1dx^2 N_c\xi  B'_3\ .}
	
	As before, we restrict to configurations with positive magnetic flux. As in \eqref{flux}, the flux is related via the Bogomol'nyi equations to the angular dependence of the $N_c$ scalars obtaining vacuum expectation values. Consider for example the asymptotic configuration
	\eql{manyphases}{\lim_{r\to\infty}\textbf{q}=\mat{ve^{ik_1\phi}&0&0&\cdots&0&0&\cdots&0\\0&ve^{ik_2\phi}&0&\cdots&0&0&\cdots&0\\0&0&ve^{ik_3\phi}&\cdots&0&0&\cdots&
			0\\\vdots&\vdots&\vdots&\ddots&\vdots&\vdots&\ddots&\vdots\\0&0&0&\cdots&ve^{ik_{N_c}\phi}&0&\cdots&0}\ ,}
	where $0\leq k_i\in\mathbb{Z}$.
	The U(1) flux can be easily calculated using the equation $(\cD_1+i\cD_2)\textbf{q}=0$,
	\eql{fluxphase}{\Phi=\lim_{r\to\infty}\int A'_\phi d\phi=\frac{2\pi\sum_{i=1}^{N_c}k_i}{C}\ .}
	
	\subsection{Size-Modes of the Minimal String ($K=1$)}\label{subsec:k=1}
	We now focus on the minimal string with topological charge $K\equiv\sum_{a=1}^{N_c}k_a=1$. We will start by constructing the basic string configuration in which only the $N_c$ scalars that acquire vacuum expectation values and the Cartan gauge fields participate. As before,  we guarantee that the terms that include $a$ and $a'$ in (\ref{tension-general}) vanish by setting the gauge multiplet scalars to their vacuum expectation values as in \eqref{generalvac}. We are then left with the Bogomol'nyi equations
	\eq{B'_3+\frac{e^2}{2}\left(\sum_i c_i|q_i|^2-N_c\xi\right)=B_3^\al+ g^2\sum_i q_i^\dagger\la^\al q_i=(\cD_1+i\cD_2)\textbf{q}=0\ .}

	We assume an ansatz of the form 
	\eql{generalbasic}{\textbf{q}=\mat{q'_{(1)}(r)&0&0&\cdots&0&0&\cdots&0\\0&q'_{(2)}(r)&0&\cdots&0&0&\cdots&0\\0&0&q'_{(3)}(r)&\cdots&0&0&\cdots&
			0\\\vdots&\vdots&\vdots&\ddots&\vdots&\vdots&\ddots&\vdots\\0&0&0&\cdots&q_v(r)e^{i\phi}&0&\cdots&0}\;,}where $q'_{(i)}(r)$  and $q_v(r)$ approach $v$ when $r\to\infty$.  $q'_{(i)}(r)$ is finite everywhere while $q_v(r)$  has a single zero at the origin.  One can show that if $c_i\neq c_j$ then $q'_{(i)}(r)\neq q'_{(j)}(r)$.\footnote{Let us choose a basis for the Cartan subalgebra of  $su(N_c)$, $\lambda^\alpha$, $\alpha=1,...,{N_c-1}$, such that
		$\lambda^{\al^*}=\text{diag}(1,-1,0,0,...,0)$ and  $\lambda^\al_{11}=\lambda^\al_{22}$ for $\al\neq {\al^*}$.
		The Bogomol'nyi equations imply that 
		\begin{equation*}
		{B_3^{\al^*}=-g^2\left(|q'_{(1)}|^2-|q'_{(2)}|^2\right)}\;,\;\;\;
		{\bar{\partial}\log\left(q'_{(1)}/q'_{(2)}\right)=i(c_1-c_2){\bar A'}+2i{\bar A}^{\al^*}}\;.
		\end{equation*}
		If $q'_{(1)}(r)=q'_{(2)}(r)$,  the first equation implies that $B_3^{\al^*}=0$ and the second equation implies that $B_3^{\al^*}= (c_2-c_1)B_3'$. Since $B_3'\neq 0$ we conclude that $c_1=c_2$. } 
	
	Let us now move on to the string zero-modes. In addition to the translation moduli we have approximate zero-modes in the $\mu_i\ll ev,gv$ limit.   For example, if the charge of the winding scalar is equal to some of the charges of the other scalars that obtain a vacuum expectation value, we have orientation zero-modes, similar to the ones discussed in the previous section. As will be described in \citep{GGKK}, there are evidences for the existence of additional zero-modes appearing in the first  $N_c^2$ hypermultiplets. We will ignore these for now as they have no effect on the question of bulk-string decoupling.

	We will focus on the size moduli, that is, on deformations of the string configuration in which  the extra $N_f-N_c$ flavors participate in the solution. As illustrated in section \ref{u1twoflav}, the decay rate of these modes will control their coupling to the bulk at low energies. 
	
	We will assume an ansatz of the form \eqref{generalbasic} for $i\leq N_c$, but allow the $\bf q$ and  $\bf \tilde q$ entries for $i>N_c$ to be non-zero. The Bogomol'nyi equations now read
	\begin{align}
	&B'_3+ \frac{e^2}{2}\left(\sum_ic_i|q_i|^2-\sum_ic_i|\tilde{q}_i|^2-N_c\xi\right)=0\;,\label{513}\\
	&B_3^\al+ g^2\sum_iq^{i\dagger}\la^\al q_i- g^2\sum_i\tilde{q}^i\la^\al\tilde{q}_i^\dagger
	=0\;,\label{514}\\&\label{barDq=0}(\cD_1+ i\cD_2)q_i=(\cD_1+ i\cD_2)\tilde{q}^i=0\;,\\&\label{F}\sum_i\tilde{q}^i\la^\al q_i=\sum_ic_i\tilde{q}^i q_i=0\;.
	\end{align}  
	
	Equation (\ref{barDq=0}) implies that 
	\begin{equation}\label{smmsms}\begin{aligned} &q_{i>N_c}^{a}=q_{a}^a\left(\prod_{b=1}^{N_c}q^b_b\right)^{\frac{c_i-c_a}{C}}f_{ia}(z)\;,\\
	&\tilde q_a^{i>N_c}=\left(q_{a}^a\right)^{-1}\left(\prod_{b=1}^{N_c}q^b_b\right)^{-\frac{c_i-c_a}{C}}\tilde f_{ia}(z)\;,\end{aligned}\end{equation}
	where $f_{ia}(z)$ and $\tilde f_{ia}(z)$ are independent of $\bar z$.
	Using  $q_b^b=q'_{(b)}(r)$ for $b<N_c$ and $q_{N_c}^{N_c}=q_v(r)e^{i\phi}$ we can write
	\begin{align}
	&q_{i>N_c}^{a<N_c}\label{11}=q'_{(a)}(r)\left(\prod_{j=1}^{N_c-1}q'_{(j)}(r)\right)^{\Delta_{ia}}\left(q_v(r)e^{i\phi}\right)^{\Delta_{ia}}f_{ia}(z)\ ,\;\;\; \Delta_{ia}\equiv \frac{c_i-c_a}{C}\;,\\
	&q_{i>N_c}^{N_c}\label{22}=\left(\prod_{j=1}^{N_c-1}q'_{(j)}(r)\right)^{\Delta_i}\left(q_v(r)e^{i\phi}\right)^{1+\Delta_i}f_i(z)\ ,\;\;\;\;\;\;\;\;\;\;\;\;\;\; \Delta_i\equiv \frac{c_i-c_{N_c}}{C}\;,\\
	&\tilde{q}_{N_c}^{i>N_c}\label{33}=\left(\prod_{j=1}^{N_c-1}q'_{(j)}(r)\right)^{-\Delta_i}\left(q_v(r)e^{i\phi}\right)^{-1-\Delta_i}\tilde f_i(z)\;,\\
	&\tilde{q}^{i>N_c}_{a<N_c}\label{44}=\onov{q'_{(a)}(r)}\left(\prod_{j=1}^{N_c-1}q'_{(j)}(r)\right)^{-\Delta_{ia}}\left(q_v(r)e^{i\phi}\right)^{-\Delta_{ia}}\tilde f_{ia}(z)\; .
	\end{align}
	
	Requiring that $q_{i>N_c}^a$ and $\tilde q^{i>N_c}_a$ are everywhere regular and decay at infinity, we can  constrain the $z$ dependence, 
	\begin{align}
	\label{1}&f_i(z)=\sum_{n=0}^{\lceil\Delta_i\rceil}\rho_i^{(n)} z^{n-1-\Delta_i}\; ,\\ 
	&f_{ia}(z)=\sum_{n=0}^{\lceil\Delta_{ia}\rceil-1}\rho_{ia}^{(n)} z^{n-\Delta_{ia}}\; ,\\&\tilde{f}_{i}(z)=\sum_{n=0}^{\lceil-\Delta_{i}\rceil-2}\tilde{\rho}_i^{(n)} z^{n+\Delta_{i}+1}\; ,\\ 
	&\label{4}\tilde{f}_{ia}(z)=\sum_{n=0}^{\lceil-\Delta_{ia}\rceil-1}\tilde{\rho}_{ia}^{(n)} z^{n+\Delta_{ia}}\; , 
	\end{align}
	where $\rho_i^{(n)},\rho_{ia}^{(n)},  \tilde{\rho}_i^{(n)},\tilde{\rho}_{ia}^{(n)}\in\mathbb{C}$. The relation between these modes and the size of the string is discussed in appendix \ref{sizeofstring}. As in the previous sections, to get the conclusions (\ref{1}-\ref{4}) we used the behavior of $q'_{(i)}(r)$ and $q_v(r)$ at $r\to\infty$ and the fact that these functions are everywhere finite, except for a single zero of the latter at the origin, where it behaves as $q_v(r)\sim r$.
	
	As a concrete example, in the theory described in section \ref{sec:HT}, $\Delta_i=\Delta_{ia}=0$ and therefore  $f_i(z)=\frac{\rho_i}{z}\ ,\ f_{ia}(z)=\tilde f_i(z)=\tilde f_{ia}(z)=0$.

	\subsection{Size-Modes of the $K$-String}
\label{Kstring}
	Let us now move on to the $K$-string.  
	As for the minimal string, we will focus here on a specific type of zero-modes -- we will assume a solution of the form \eq{q_{i\leq N_c}^a=\delta_{i}^aq_a(r,\phi)\;,\;\;\; \tilde q^{i\leq N_c}_a=0\;,} 
	and study the freedom in the $i>N_c$ entries. The free parameters in $q_{i>N_c}^a$ and $\tilde q^{i> N_c}_a$ will be referred to as the size-modes.

	The boundary conditions are given by $\lim\limits_{r\to\infty}q_a(r,\phi)=ve^{ik_a\phi}$, $0\leq k_a\in\mathbb{Z}$, $\sum_{a=1}^{N_c}k_a=K$.  Together with the Bogomol'nyi equations this implies that $q_a(r,\phi)$ has $k_a$, not necessarily distinct, zeros  at positions $z_{l_a}$. Close to the zeros it behaves as $q_a\sim z-z_{l_a}$. 
	
	As  before, equation (\ref{barDq=0}) implies \eqref{smmsms}. Using the boundary conditions we get that the size-modes take the form
	\eql{sizemodesgeneralzeros}{q_{i>N_c}^{a}&=\left(\prod_{b\neq a}\frac{q_{b}(r,\phi)}{\prod_{l_b=1}^{k_b}(z-z_{l_b})}\right)^{\Delta_{ia}}\left(\frac{q_a(r,\phi)}{\prod_{l_a=1}^{k_a}(z-z_{l_a})}\right)^{1+\Delta_{ia}}\times\sum_{n=0}^{\lceil k_a+\Delta_{ia} K\rceil-1}\rho^{(n)}_{ia}z^n\;,\\
		\tilde{q}^{i>N_c}_{a}&=\left(\prod_{b\neq a}\frac{q_{b}(r,\phi)}{\prod_{l_b=1}^{k_b}(z-z_{l_b})}\right)^{-\Delta_{ia}}\left(\frac{q_a(r,\phi)}{\prod_{l_a=1}^{k_a}(z-z_{l_a})}\right)^{-\Delta_{ia}-1}\times \sum_{n=0}^{\lceil-k_a-\Delta_{ia}K\rceil-1}\tilde{\rho}^{(n)}_{ia}z^n\;,}
	with ${\rho}_{ia}^{(n)},\tilde{\rho}_{ia}^{(n)}\in\mathbb{C}$.

	For simplicity, we will assume in the analysis below that  all the zeros coincide at the origin. Equation \eqref{sizemodesgeneralzeros} then becomes
	\eql{sizemodesatorigin}{q_{i>N_c}^{a}&=\left(\prod_{b\neq a}q_{b}(r,\phi)\right)^{\Delta_{ia}}\left(q_a(r,\phi)\right)^{1+\Delta_{ia}}\sum_{n=0}^{\lceil k_a+\Delta_{ia} K\rceil-1}\rho^{(n)}_{ia}z^{n-\Delta_{ia}K-k_a}\;,\\
		\tilde{q}^{i>N_c}_{a}&=\left(\prod_{b\neq a}q_{b}(r,\phi)\right)^{-\Delta_{ia}}\left(q_a(r,\phi)\right)^{-\Delta_{ia}-1}\sum_{n=0}^{\lceil-k_a-\Delta_{ia}K\rceil-1}\tilde{\rho}^{(n)}_{ia}z^{n+\Delta_{ia}K+k_a}\;. }

	\section{Bulk-String Factorization}
	
	\label{factorization}
	
	We now discuss the factorization of the bulk and string degrees of freedom, which is the main subject of this work. 
	From equation \eqref{sizemodesatorigin} we see that  whenever $\frac{(c_i-c_a)K}{C}\slashed{\in}\mathbb{Z}$ for some $i>N_c$ and $a\leq N_c$, there exist long-range size-modes with asymptotic falloff of $\sim \onov{r^\be}\ ,\ 0<\be<1$. As explained in section \ref{u1twoflav}, such  long-range size-modes are coupled to the bulk light excitations even at low energies. In this case, the worldsheet theory cannot be separated from the bulk and the low-energy dynamics in the string background is not described in terms of a two-dimensional worldsheet theory. On the other hand, if $\frac{(c_i-c_a)K}{C}\in\mathbb{Z}$ for all $i$ and $a$, the slowest falloff is $\sim\onov{r}$. In this case, the mixing between the bulk and the string light excitations is suppressed at energy scales $\mu_i\lesssim E\ll m_W$. As a result, the low-energy effective action factorizes,
	\eql{actionfactorization}{S_{\text {eff}}=S_{\text {bulk}}+S_{\text {worldsheet}}\ .}

	The relation between the asymptotic behavior of the size-modes and the bulk-string factorization was already explained in section \ref{u1twoflav}. For completeness, we will summarize the main steps here, this time in the context of   the theory described in section \ref{general gauging}.

	As explained before, in the presence of non-degenerate masses there are no true zero-modes, except for the center-of-mass modes.  We wish, however, to distinguish between heavy modes with mass  of order $m_W$ and light modes with mass of the order of the hypermultiplet masses. Since we are interested in fluctuations with energies $\mu_i\lesssim E\ll m_W$, we can ignore fluctuations of the heavy modes and treat such modes as background fields with their value fixed by their vacuum expectation value.
	
	The heavy modes are the $N_c^2$ long vector multiplets.   The light modes are the $N_c(N_f-N_c)$ hypermultiplets labeled by $i>N_c$, as well as  the bosonic (approximate) moduli discussed in the previous section and their superpartners. 
	In terms of the light modes, the scalars $q_{i>N_c}^a$, $\tilde q^{i>N_c}_a$ can be written as 
	\eql{sizeexpansion}{q=q_{\text{bulk}}+\sum_nq_{\text{size}}^{(n)}\;,}
	where $q_{\text{bulk}}$ are the bulk excitations  and $q_{\text{size}}^{(n)}$ are the different terms that appear in the expansion  \eqref{sizemodesatorigin}. Asymptotically, $|q_{\text{size}}^{(n)}|\sim r^{-\be_n}$, where \eq{\be_n=\begin{cases} -n+\Delta_{ia}K+k_a \;\text{ for }q_i^a\\ -n-\Delta_{ia}K-k_a\;\text{ for }\tilde{q}_a^i\end{cases}\ .}
	The distinction between the bulk and string  excitations makes sense only if the string and the bulk modes decouple at energies $\mu_i\lesssim E\ll m_W$. We assume that this is the case and obtain a contradiction when $\be<1$ modes exist.

	We can repeat the computation of section \ref{u1twoflav} and canonically normalize the size-modes as fields on the worldsheet, see the discussion around equations (\ref{kinint}-\ref{canondef}). Then, writing the action in terms of the canonically normalized fields we can evaluate the operators that mix the size-modes and the bulk excitations.  Terms that are suppressed in the $\mu_i\lesssim E\ll m_W$ limit are thrown away. 
	
	The mixing coming from the kinetic terms, for example, has been studied in section ֿ\ref{u1twoflav} (see the discussion after equation \eqref{mixedkinetic}) where we found that this term is irrelevant at energies $E\ll m_W$ if $\be>1$. For $\be=1$ the coupling is classically marginal, but suppressed in the $\mu_i\ll m_W$ limit. For $\be<1$ the coupling is important at low energies, $E\sim \mu_i$, thus contradicting the decoupling of $\be<1$ modes from the bulk. It is not difficult to check that for $\be\geq1$ the other mixing terms are also suppressed at energies $\mu_i\lesssim E\ll m_W$. 
	
	Thus, the result of this classical analysis is that the bulk and the string modes decouple if and only if $\frac{(c_i-c_a)K}{C}\in\mathbb{Z}$ for all $a\leq N_c<i$. 
	One may doubt the validity of this conclusion in the quantum theory, as our analysis relies heavily  on naive classical arguments. However, exactly the same conclusions are obtained from a non-perturbative localization computation, as we explain in section \ref{localization}. 

We now show that the decoupling criterion above coincides with the condition of trivial Aharonov-Bohm phases.  Using the Bogomol'nyi equation $(\cD_1+i\cD_2)\textbf{q}=0$ and the asymptotic behavior we obtain
\eq{\lim_{r\to\infty}\int d\phi\left[c_aA'_\phi+\sum_{\al\in \text{Cartan}}\la^\al_{aa}A^\al_\phi\right]= 2\pi k_a\ ,\;\;\;a=1,...,N_c\;.}
Using this we can compute the phase acquired by a $q_i^a$-particle when circling the string at $r\to\infty$,
\eq{\lim_{r\to\infty}\int d\phi\left[c_iA'_\phi+\sum_{\al\in \text{Cartan}}\la^\al_{aa}A^\al_\phi\right]=\frac{2\pi(c_i-c_a)K}{C}+2\pi k_a\ .}
We thus find that \eqref{decoupling criterion} is exactly the condition that the phases acquired by the light particles in the bulk upon circling the string are trivial.

\section{Merging Size-Modes}
\label{Kstringmoduli}
As was shown in the previous sections, whenever $\frac{(c_i-c_a)K}{C}\slashed{\in}\mathbb{Z}$ for some $a\leq N_c<i$, the $K$-string is coupled to the bulk due to the existence of long-range size-modes. When ${\frac{(c_i-c_a)K}{C}\in\mathbb{Z}}$ for all $a\leq N_c$ and $i>N_c$, the size-modes decay at least as fast as $1/{r}$, thus guaranteeing  the decoupling between the bulk and the string. We see that the same type of string can be decoupled or coupled to the bulk, depending on its topological charge $K$. In this section, we discuss some aspects of the decoupled $K$-string, in cases where the minimal string with $K=1$ is coupled to the bulk.

When the minimal string is decoupled from the bulk, no long-range modes exist and we can use our usual intuition for BPS objects. We can imagine taking the $K$-string with topological charge $K$ and separating it  into $K$ minimal strings. At large separation, the moduli space should factorize into $K$ copies of the moduli space of the minimal string.  In particular, one expects the dimension of the $K$-string moduli space to be $K$ times the dimension of the minimal string moduli space.

This intuition fails if the minimal string is coupled to the bulk. In this case, the $K$-string cannot be separated into $K$ minimal strings; we still expect that the $K$-string can be separated spatially into a configuration with $K$ distinct zeros, but these zeros cannot be thought of as isolated minimal strings. We suggest that the long-range size-modes around the zeros are combined into fewer modes with faster decay rates. These modes are not associated with one minimal string. They are collective modes related to the $K$-string as a whole. In particular, the dimension of the $K$-string moduli space is not necessarily an integer multiple of $K$ and is expected to be smaller than $K$ times the dimension of the minimal string moduli space.

Evidence in favor of this interpretation is obtained by counting the number of the $K$-string size-modes using equation (\ref{sizemodesgeneralzeros}). Let us assume for simplicity that $\Delta_{ia}\geq0$ for every $a\leq N_c$ for some $i>N_c$, such that the flavor $q_i$ gives rise to $\sum_{a=1}^{N_c}\left(k_a+\Delta_{ia}K\right)=\frac{c_i N_c K}{C}$ complex size-modes. This number is not necessarily an integer multiple of $K$. Counting the number of size-modes for the minimal string using (\ref{sizemodesgeneralzeros})  we find $1+\sum_{a=1}^{N_c}\lceil\Delta_{ia}\rceil$ size-modes related to $q_i$. Thus,  the number of the $K$-string size-modes is equal to $K$ times the number of the minimal string size-modes if and only if the minimal string is decoupled from the bulk. This is in agreement with the expectation based on separating the  $K$-string into $K$,  far apart,  minimal strings. On the other hand, if the minimal string is coupled to the bulk, the number of size-modes of the $K$-string is smaller than   $K$ times the number of size-modes of the minimal string. This supports the suggestion that the long-range size-modes are combined into fewer modes with faster decay rates.

It is important to emphasize that the discussion above does not apply for the center-of-mass and orientation modes, whose number is $K$ times the number of the corresponding modes for the minimal string. This must be true because we can, in principle, take the mass of the $i>N_c$ flavors to be very large such that the extra flavors decouple.   The number of center-of-mass and orientational modes shouldn't be affected by this decoupling.

\section{Supersymmetric Localization Derivation of the Decoupling Criterion}
\label{localization}

In recent years, many exact results have been obtained for BPS objects in supersymmetric theories on spheres, using the technique of supersymmetric localization. In particular, exact (non-perturbative) formulas are available for 
squashed sphere partition functions of  four-dimensional $\cN=2$ Lagrangian theories that preserve a $U(1)\subset SU(2)_R$ symmetry \cite{Pestun:2007rz,Hama:2012bg}, as well as  for two-dimensional $\cN=(2,2)$ theories with a $U(1)$ $R$-symmetry \cite{Doroud:2012xw,Benini:2012ui}.  In an upcoming publication we use this fact to study the worldsheet theories of the strings presented in section~\ref{general gauging}. In this section we will briefly review how the decoupling criterion (\ref{decoupling criterion}) arises in the localization analysis that will appear in \citep{GGKK}. 

The $\cN=2$ supersymmetric theory described in section \ref{general gauging}  can be placed on the four-ellipsoid,  
\begin{equation}
\frac{x_0^2}{r^2}+\frac{x_1^2+x_2^2}{l^2}+\frac{x_3^2+x_4^2}{\tilde l^2}=1\;,\label{ellipsoid}
\end{equation} while preserving a supercharge that squares to a combination of rotations and an $R$-symmetry transformation \cite{Hama:2012bg}. 

Using the localization formula of \cite{Hama:2012bg}, the four-ellipsoid partition function can be formally expressed as a matrix integral over the Coulomb branch coordinates. This formal expression is, however, divergent due to the $U(1)$ Landau-pole. To extract information from this expression we introduce an ultraviolet  cut-off, and cut-off  the Coulomb-branch integral at a scale $\Lambda$, much smaller than the scale set by the Landau-pole.  

In analogy to our assumptions in the previous sections, we assume that $\mu_i,l^{-1},\tilde{l}^{-1}\ll \xi g^2,\xi e^2$.\footnote{In this section we follow the notations of \cite{Hama:2012bg} for the Fayet-Iliopoulos parameter,  in which  $\xi$ has mass dimension $1$. In the case of the round sphere, this  parameter is related to the one we have in flat space by rescaling with the radius. } Then, for a wide range of $\Lambda$-values, we can close the Coulomb-branch integrals in the complex plane, and compute the integrals using Cauchy's theorem. In the spirit of \cite{Doroud:2012xw,Benini:2012ui}, we obtain a representation of the four-ellipsoid partition function as a discrete sum, which we interpret as a sum over string contributions.\footnote{The interpretation of the residues of the poles of the integrand as encoding the string contributions is analogous to the analysis of \cite{Gaiotto:2012xa}, in which singularities of the superconformal index have been identified with surface defects. } Indeed, in \cite{Chen:2015fta,
	Pan:2015hza} it was shown that with a modified deformation term, the path integral localizes on saddle points that solve a BPS string-like equation. This modified localization prescription results in a Higgs branch representation of the partition function. It was claimed in  \cite{Chen:2015fta,
	Pan:2015hza}  that this representation  can be obtained from the Coulomb-branch representation of \cite{Pestun:2007rz,Hama:2012bg} upon closing the contours in the complex plane.

In \citep{GGKK}, we follow the path described above and obtain a representation of the four-ellipsoid partition function of the form 
\begin{equation}
\begin{aligned}
\sum_{\{l_a\}}Z^{B}_{\text{vac},\{l_a\}}e^{-8\pi^2\hx{N_c(b+b^{-1})}/|C_{\{l_a\}}|}e^{-16i\pi^2\hx{\sum_{a=1}^{N_c}\hm_{l_a}}/C_{\{l_a\}}}
\sum_{K,K'}e^{-16\pi^2\hx( K b+ K' b^{-1})/|C_{\{l_a\}}|}\,Z^{B}_{K,K'\{l_a\}}+...\label{expression}
\end{aligned}
\end{equation}
In this expression the sum over ${\{l_a|a=1,...,N_c\}\subset\{1,2,...,N_f\}}$ is a sum over choices of $N_c$ out of the $N_f$ hypermultiplets for which $C_{\{l_a\}}\equiv\sum_{a=1}^{N_c}c_{l_a}\neq 0$ -- this is interpreted as a sum over the baryonic vacua. 	The dots stand for a sum over mesonic vacua, which we do not present here.  We have denoted by $b$ the squashing parameter,  $b\equiv\sqrt{{l}/{\tilde l}}$, and used a hat for dimensionless quantities, rescaled as $\hm=\sqrt{l\tilde l}\mu$, $\hx=\sqrt{l\tilde l}\xi$. 

The expression in (\ref{expression}) has been expanded according to the dependence on the  Fayet-Iliopoulos parameter $\xi$. This expansion defines the functions $Z^{B}_{\text{vac},\{l_a\}}$ and $Z^B_{K,K',\{l_a\}}$,  which are functions of the rescaled hypermultiplet masses $\hm_i$, the gauge couplings, the theta parameters, and the squashing parameter $b$.  We now explain our interpretation of the different terms in \eqref{expression}.   

The function $Z^{B}_{\text{vac},\{l_a\}}$ is identified with  the four-ellipsoid partition function describing the  excitations around the vacuum labeled by $\{l_a\}$. Indeed, this factor is equal to the four-ellipsoid partition function of the $N_c(N_f-N_c)$ light  hypermultiplets in this vacuum. 

$K$ and $K'$ label the winding numbers transverse to the squashed two-spheres defined by ${x_0^2}/{r^2}+\left({x_1^2+x_2^2}\right)/{l^2}=1$ and ${x_0^2}/{r^2}+\left({x_3^2+x_4^2}\right)/{\tilde l^2}=1$ respectively. 
The  functions $Z^B_{K,K',\{l_a\}}$  are weighted in this sum with an $e^{-16\pi^2\hx( K b+ K' b^{-1})/|C_{\{l_a\}}|}$ factor, related to the action of the string configuration. The $U(1)$ fluxes through the two two-spheres can be read from this factor. We use this factor to separate contributions of strings with different winding numbers and to separate contributions of strings wrapping different  two-spheres.  We focus on contributions with $K'=0$ --  these correspond to strings with winding number $K$, whose cores wrap the two-sphere ${x_0^2}/{r^2}+\left({x_1^2+x_2^2}\right)/{l^2}=1$.\footnote{Contributions for which both $K$ and $K'$ are not zero correspond to a $K$-string and a $K'$-string, wrapping the two-spheres $\frac{x_0^2}{r^2}+\frac{x_1^2+x_2^2}{l^2}=1$ and $\frac{x_0^2}{r^2}+\frac{x_3^2+x_4^2}{\tilde l^2}=1$ respectively, and intersecting in the two-poles $x_0=\pm r$.  See \cite{Chen:2015fta,
		Pan:2015hza,Pan:2016fbl}. } 

In \cite{Chen:2015fta} (see also \cite{Pan:2015hza,Pan:2016fbl,Gomis:2016ljm}), an expression similar to (\ref{expression}) was derived for the $U(N_c)$ gauge theory described in section \ref{sec:HT}.  In this case, $Z^B_{K,\{l_a\}}\equiv Z^B_{K,K'=0,\{l_a\}}$ was identified with the two-sphere partition function of the worldsheet theory proposed in \cite{Hanany:2003hp}.  Other examples in which $Z^B_{K,\{l_a\}}$ is identified with a two-sphere partition function of an $\cN=(2,2)$ sigma-model will be presented in \citep{GGKK}. In these examples, we claim that the sigma models that have been identified are the corresponding worldsheet theories, and provide consistency checks supporting these claims.

The factorization of the string contributions in (\ref{expression}) to a product of a four-ellipsoid partition function describing the light fields in the bulk and a two-sphere partition function describing the interactions of the string moduli is interpreted as a sign of the decoupling of the string moduli from the bulk modes. This factorization applies  only for strings that satisfy the decoupling condition (\ref{decoupling criterion}), as we explain below.  

The one-loop determinant of the localization formula of \cite{Hama:2012bg} is given in terms of $\Upsilon_b$-functions. The function $\Upsilon_b(x)$ is a holomorphic function, invariant under $b\to b^{-1}$, which is uniquely defined by the normalization $\Upsilon_b\left(\half(b+b^{-1})\right)=1$ and the shift relation \begin{equation}
\Upsilon_b\left(x+b\right)=\frac{\Gamma\left(bx\right)}{\Gamma\left(1-bx\right)}b^{1-2bx}\Upsilon_b\left(x\right)\;.\label{shift}
\end{equation} For any choice of a baryonic vacuum and a winding number, we obtain from the expansion (\ref{expression}) an expression for $Z^B_{K,\{l_a\}}$, which contains  ratios of $\Upsilon_b$-functions. 
When  (\ref{decoupling criterion}) is satisfied, the differences between the arguments of the numerator and denominator $\Upsilon_b$-functions are integer multiples of $b$. In this case,  $Z^B_{K,\{l_a\}}$ can  be rewritten in terms of Gamma functions using the shift relation (\ref{shift}).\footnote{If both $K$ and $K'$ are not zero, one has to demand (\ref{decoupling criterion}) for both $K$ and $K'$. } Doing this, we   find expressions for $Z^B_{K,\{l_a\}}$ of the form obtained in \cite{Doroud:2012xw,Benini:2012ui} for two-sphere partition functions in the Higgs-branch representation. On the other hand, when (\ref{decoupling criterion}) is not satisfied and one cannot use (\ref{shift}) to rewrite $Z^B_{K,\{l_a\}}$ in terms of Gamma functions, an interpretation of $Z^B_{K,\{l_a\}}$ as a two-sphere partition function seems unlikely. 

We thus interpret the condition (\ref{decoupling criterion}) as a criterion for the decoupling of the string from the bulk. 
Remarkably, this criterion coincides exactly with the one we obtained  using semiclassical methods in section \ref{factorization}.

\section*{Acknowledgments}

We would like to thank O. Mamroud, D.~Tong and T.~Vaknin for fruitful discussions. We especially thank J. Gomis and Z. Komargodski for collaboration throughout a
large portion of this project and for many useful discussions. 
We thank the Galileo Galilei Institute for Theoretical Physics and the Perimeter
Institute for Theoretical Physics for hospitality during the course of this project. This research was supported in part by Perimeter Institute for Theoretical Physics. Research at Perimeter Institute is supported by the
Government of Canada through the Department of Innovation, Science and Economic Development and by the Province of Ontario through
the Ministry of Research and Innovation.
E.G. and A.K. are supported by the ERC STG grant 335182.

\appendix
\section{The Size of the String}
\label{sizeofstring}
In this appendix we will find the asymptotic behavior of the string configuration in the presence of size-modes. While in the absence of size-modes the string decays exponentially fast to the vacuum, when  the size-modes are present the string decays to the vacuum with a power-law behavior governed by the modes with the slowest decay. The typical size of the string is related to the absolute value of the size modulus $|\rho|$ of the slowest decaying mode.

We will start from the $U(1)$ theory with two charged flavors of section \ref{u1twoflav}. 
Plugging equation \eqref{q2zero} into the Bogomol'nyi equations \eqref{Bogmol'nyi-u(1)mu0} we obtain
\eql{originalbog}{\bar{A}=-\frac{i}{p}\bar{\partial}\log q_1\ ,\;\;\; \frac{2B_3}{e^2}+ p|q_1|^2+|q_1|^{2/p}\left|\sum_{n=0}^{\lceil k/p\rceil-1} \frac{\rho^{(n)}}{z^{k/p-n}} \right|^2=\xi\ .}
Using $B_3=2i(\bar{\partial}A-\partial\bar{A})$ we can eliminate $A_\mu$ from the equations. We are left with one equation for $q_1$,
\eql{eqqq}{\frac{1}{pe^2}\left[\onov{r}\dr\left(r\dr\log|q_1|^2\right)+\onov{r^2}\partial_\phi^2\log|q_1|^2\right]= p|q_1|^2+|q_1|^{2/p}\left|\sum_{n=0}^{\lceil k/p\rceil-1} \frac{\rho^{(n)}}{z^{k/p-n}} \right|^2-\xi\ .}

As we are only interested in the asymptotic behavior, we can keep in the sum over size-modes only the leading term at large $r$ and replace
\eql{leadingsize}{\left|\sum_{n=0}^{\lceil k/p\rceil-1} \frac{\rho^{(n)}}{z^{k/p-n}} \right|^2\rightarrow\; \frac{|\rho_{\text{min}}|^2}{r^{2\be_{\text{min}}}}\; ,}
where $\rho_{\text{min}}=\rho^{(\lceil k/p\rceil-1)}$, $\be_{\text{min}}=k/p-\lceil k/p\rceil+1$.\footnote{We assume that $\rho_{\text{min}}\neq 0$. Otherwise, one has to take the slowest decaying mode for which $\rho\neq 0 $.}

Defining   $|q_1|^2=\frac{\xi}{p}(1-X)$, and remembering that $\lim_{r\to\infty}X=0$, we can linearize (\ref{eqqq}) in the large $r$ limit, 
\eql{a4}{\onov{pe^2}\left[\frac{1}{r}{\dr X}+\dr^2 X\right]=\xi X-\left(\frac{\xi}{p}\right)^{\onov{p}} \frac{|\rho_{\text{min}}|^2}{r^{2\be_{\text{min}}}}\ .}
The homogeneous equation is solved by
\begin{equation}
X_h\sim e^{-\sqrt{pe^2\xi}r}\left(1+ֿ\cO\left(\onov{e\sqrt{\xi} r}\right)\right)\;.  
\end{equation}
Thus, in the absence of size-modes, the string solution has an exponential falloff, with typical size $(pe^2\xi)^{-{1}/{2}}$. 

For non-zero $\rho^{(n)}$, equation (\ref{a4}) is solved asymptotically by  
\eql{privatesolution}{X=\onov{\xi}\left(\frac{\xi}{p}\right)^{\onov{p}} \frac{\left|\rho_{\text{min}}\right|^2}{r^{2\be_{\text{min}}}}\left(1+\cO\left(\onov{e^2{\xi} r^2}\right)\right)\ .}
Thus, the string has a power-law falloff at large distances with a typical size 
\eql{rstorho}{r_s=\left(\frac{\xi^{\frac{1-p}{p}}}{p^{1/p}}|\rho_{\text{min}}|^2\right)^{\onov{2\be_{\text{min}}}}\ .}

The generalization to strings in $SU(N_c)\times U(1)$ gauge theories is straightforward. For example, take the string solution of the type studied in sections \ref{subsec:k=1}-\ref{Kstring},  and assume that the slowest decaying size-mode belongs to the $q_i^a$ component. Using equation (ֿֿ\ref{sizemodesatorigin}), we will denote \eq{\be_{\text{min}}=\Delta_{ia}K-\lceil\Delta_{ia}K\rceil+1\ ,\;\;\; \rho_{\text{min}}=\rho_{ia}^{(k_a+\lceil\Delta_{ia}K\rceil-1)}\ ,}such that \eq{\ \lim_{r\to\infty}|q_i^a|^2=\frac{v^{2(\Delta_{ia}N_c+1)}|\rho_{\text{min}}|^2}{r^{2\be_{\text{min}}}}\;.} Plugging equations (\ref{513}-\ref{514}) into equation \eqref{barDq=0}, and linearizing the resulting equation, we obtain a set of $N_c$ coupled asymptotic equations,
\eq{\frac{1}{2r}\dr X_b+\half \dr^2 X_b- g^2v^2\lambda^\alpha_{bb}\sum_{c=1}^{N_c}\lambda^\alpha_{cc}X_c-\frac{e^2v^2}{2}c_b\sum_{c=1}^{N_c}c_cX_c=-\left(\frac{e^2v^2}{2}c_bc_i+g^2v^2\lambda^\alpha_{bb}\lambda^\alpha_{aa}\right)v^{2\Delta_{ia}N_c}\frac{|\rho_{\text{min}}|^2}{r^{2\be_{\text{min}}}}\;,}
for $b=1,..,N_c$, where we denoted $|q_b^b|^2=v^2\left(1-X_b\right)$. 

In the absence of size-modes, we need to consider the solutions to the homogeneous matrix equation,
 \eq{\frac{1}{r}\dr X_b+ \dr^2 X_b=M_{bc}X_c\ ,\;\;\;\ M_{bc}= 2g^2v^2\lambda^\alpha_{bb}\lambda^\alpha_{cc}+e^2v^2c_bc_c\;.}
Asymptotically, the one derivative term is negligible, and the $r$ dependence of the solutions is of the form,
 \eq{X_b=\sum_{I=1}^{N_c}X_b^{(I)}e^{-r/r_I}}
where $r_I^{-2}$ are the eigenvalues of $M_{bc}$. While the exact form of $r_I$ is complicated,  it is easy to see that $r_I$ is proportional to $v^{-1}$ and goes to zero in the large $m_W$ limit.

Turning on the size-mode, the derivative terms are negligible at large distances and the asymptotic behavior is controlled by the set of algebraic equations 
\eq{M_{bc}X_c=\left(e^2c_bc_i+2g^2\lambda^\alpha_{bb}\lambda^\alpha_{aa}\right)v^{2\Delta_{ia}N_c+2}\frac{|\rho_{\text{min}}|^2}{r^{2\be_{\text{min}}}}\;.}
These equations are solved by $X_b=\left(\frac{r_b}{r}\right)^{2\be_{\text{min}}}$ with $r_b$ proportional to $\left(v^{\Delta_{ia}N_c}{|\rho_{\text{min}}|}\right)^\frac{1}{\be_{\text{min}}}$.

\bibliography{FactorizationFinal}

	\end{document}